\documentclass[usenatbib]{mnras}
\usepackage{graphicx}
\usepackage{amsmath}
\usepackage{breqn}
\usepackage{amssymb}
\usepackage{hyperref}
\usepackage{xcolor}

\title[$z\sim6$ quasars: Impact of black hole seeding models]{Probing the $z\gtrsim6$ quasars in a universe with IllustrisTNG physics: Impact of gas-based black hole seeding models}
\author[Bhowmick et al.]{Aklant K. Bhowmick$^{1}$,
Laura Blecha$^{1}$,
Yueying Ni$^{2,3}$,
Tiziana Di Matteo$^{2,3}$,
Paul Torrey$^{1}$, \newauthor
Luke Zoltan Kelley$^{4}$, 
Mark Vogelsberger$^{5}$,
Rainer Weinberger$^{6}$,
Lars Hernquist$^{7}$
\\
$^{1}$Dept. of Physics, University of Florida, Gainesville, FL 32611, USA\\
$^{2}$McWilliams Center for Cosmology, 5000 Forbes Avenue, Pittsburgh, PA 15213, USA\\
$^{3}$ NSF AI Planning Institute for Physics of the Future, 
Carnegie Mellon  University, Pittsburgh, PA 15213, USA \\
$^{4}$Center for Interdisciplinary Exploration and Research in Astrophysics, Northwestern University, Evanston, IL 60208, United States\\
$^{5}$Dept. of Physics, Kavli Institute for Astrophysics and Space Research, Massachusetts Institute of Technology,
Cambridge, MA 02139, USA \\
$^{6}$Canadian Institute for Theoretical Astrophysics, 60 St. George Street, Toronto, ON M5S 3H8, Canada\\
$^{7}$Harvard-Smithsonian Center for Astrophysics, 60 Garden Street, Cambridge, MA 02138, USA\\
}
\begin{document}
\maketitle
\begin{abstract}
We explore the implications of a range of black hole~(BH) seeding prescriptions on the formation of the brightest $z\gtrsim6$ quasars in cosmological hydrodynamic simulations. The underlying galaxy formation model is the same as in the \texttt{IllustrisTNG} simulations. Using initial conditions generated in constrained Gaussian realizations, we study the growth of BHs in rare overdense regions~(forming $\gtrsim10^{12}~M_{\odot}/h$ halos by $z=7$) using a $~(9~\mathrm{Mpc}/h)^3$ simulated volume. BH growth is maximal within halos that are compact and have a low tidal field. For these halos, we consider an array of gas-based seeding prescriptions wherein $M_{\mathrm{seed}}=10^4-10^6~M_{\odot}/h$ seeds are inserted in halos above critical thresholds for halo mass and  dense, 
metal-poor gas mass 
(defined as $\tilde{M}_{\mathrm{h}}$ and $\tilde{M}_{\mathrm{sf,mp}}$, respectively, in units of $M_{\mathrm{seed}}$). We find that a seed model with $\tilde{M}_{\mathrm{sf,mp}}=5$ and $\tilde{M}_{\mathrm{h}}=3000$ successfully produces a $z\sim6$ quasar with $\sim10^9~M_{\odot}$ mass and  $\sim10^{47}~\mathrm{ergs~s^
{-1}}$ luminosity. BH mergers play a crucial role at $z\gtrsim9$, causing an early boost in BH mass at a time when accretion-driven BH growth is negligible. When more stringent seeding conditions are applied~(
e.g., $\tilde{M}_{\mathrm{sf,mp}}=1000$), the relative paucity of BH seeds results in a much lower merger rate.  
In this case, $z\gtrsim6$ quasars can only be formed if we enhance the maximum allowed BH accretion rates~(by factors $\gtrsim10$) compared to the accretion model used in \texttt{IllustrisTNG}. This can be achieved either by allowing for super-Eddington accretion, or by
reducing the radiative efficiency. Our results demonstrate that progenitors of $z\sim6$ quasars have distinct BH merger histories for different seeding models, which will be distinguishable with LISA observations. 

\end{abstract}

\begin{keywords}
(galaxies:) quasars: supermassive black holes; (galaxies:) formation; (galaxies:) evolution; (methods:) numerical 
\end{keywords}

\section{Introduction}

The growing population of known luminous~($\sim10^{47}~\mathrm{erg~s^{-1}}$) quasars at the highest redshifts~($z\gtrsim6$) shows that some supermassive black holes~(SMBHs) were already in place within the first $\sim1~\mathrm{Gyr}$ since the Big Bang. The inferred masses of these quasars range from $\sim10^9-10^{10}~M_{\odot}$, similar to the most massive SMBHs in the local universe. 
To date, $\gtrsim200$ quasars have already been discovered at $z\gtrsim6$~\citep{2001AJ....122.2833F,2010AJ....139..906W,2015MNRAS.453.2259V,2016ApJ...833..222J,2016Banados,2017MNRAS.468.4702R,2018ApJS..237....5M}, which overall correspond to number densities of $1~\mathrm{cGpc}^{-3}$. Additionally, there are 
a handful of objects discovered at $z\gtrsim7$~\citep{2011Natur.474..616M,2018ApJ...869L...9W,2019ApJ...872L...2M,2019AJ....157..236Y}, which includes the most distant quasars observed to date \citep{2018Natur.553..473B,2021ApJ...907L...1W} at $z\sim7.6$. The recently launched James Webb Space Telescope 
\citep[JWST;][]{2006SSRv..123..485G} and planned facilities such as Lynx X-ray Observatory  
\citep{2018arXiv180909642T} have a promising prospect of revealing the AGN~(active galactic nuclei) progenitors of these quasars at even higher redshifts. Additionally, gravitational wave events from Laser Interferometer Space Antenna  
\citep[LISA;][]{2019arXiv190706482B} will also provide insights into the prevalence of BH mergers and the growth history of these quasars. These observations are going to be crucial to understanding the assembly of these quasars, which is an outstanding challenge for theoretical models of BH formation and growth.

The origin 
of these $z\gtrsim6$ quasars, and the larger SMBH populations in general, is a subject of active debate. Remnants of the first generation of Pop III stars, a.k.a Pop III BH seeds, are popular candidates  ~\citep{2001ApJ...550..372F,2001ApJ...551L..27M,2013ApJ...773...83X,2018MNRAS.480.3762S}. The BH seed mass that results from the conjectured Pop III scenario depend on the initial mass function of Pop III stars themeselves. This is predicted to be more top heavy than present day stellar populations, with masses typically ranging from $\sim10-100~M_{\odot}$~\citep{2014ApJ...781...60H, 2016ApJ...824..119H}. But even the most massive Pop III seeds~(initial BH masses of $\sim10^2~M_{\odot}$)  would require significant periods of super-Eddington accretion to grow by $\gtrsim7$ orders of magnitude to form a $z\gtrsim6$ quasar. These stringent growth rate requirements can be alleviated to an extent with channels producing more massive seeds. Theories proposed for massive seed formation 
include runaway collisions of stars or black holes in dense nuclear star clusters forming ``NSC seeds" with masses $\sim10^{2}-10^{3}~M_{\odot}$ ~\citep{2011ApJ...740L..42D,2014MNRAS.442.3616L,2020MNRAS.498.5652K,2021MNRAS.503.1051D,2021MNRAS.tmp.1381D}, and direct collapse of gas in atomic cooling~($T_{\mathrm{vir}}>10^{4}~\mathrm{K}$) halos forming ``direct collapse black hole (DCBH) seeds" with masses $\sim10^{4}-10^{6}~M_{\odot}$~\citep{2003ApJ...596...34B,2006MNRAS.370..289B,2014ApJ...795..137R,2016MNRAS.458..233L,2018MNRAS.476.3523L,2019Natur.566...85W,2020MNRAS.492.4917L}. 

The most massive DCBH seeds are seen as promising candidates for explaining the rapid formation of the $z\gtrsim6$ quasars. Their formation requires gas to undergo an isothermal collapse at temperatures $\gtrsim10^4~\mathrm{K}$~(corresponding to a Jean's mass $\gtrsim10^{4}~M_{\odot}$). For this to occur, the gas needs to be devoid of chemical species that are efficient coolants at $\lesssim10^4~\mathrm{K}$, namely metals and molecular hydrogen. To suppress molecular hydrogen, the gas must be exposed to Lyman Werner radiation with minimum fluxes $\gtrsim1000~J_{21}$ as inferred from small scale hydrodynamic simulations~\citep{2010MNRAS.402.1249S} as well as one-zone chemistry models~\citep{2014MNRAS.445..544S,2017MNRAS.469.3329W}. Such high fluxes can only be provided by nearby star forming galaxies~\citep{2014MNRAS.445.1056V,2017NatAs...1E..75R,2021MNRAS.503.5046L,2021MNRAS.tmp.3110B}. However, these star forming regions can also pollute the gas with metals, which would then eliminate any possibility of direct collapse. Overall, this implies that the window for DCBH seed formation is extremely narrow, and it is unclear whether they form abundantly enough to explain the inferred densities of these objects. 


Semi analytic models~(SAMs) have 
so far been extensively used in the modeling of black hole seeds~\citep{2007MNRAS.377.1711S,Volonteri_2009, 2012MNRAS.423.2533B,2018MNRAS.476..407V, 2018MNRAS.481.3278R, 2019MNRAS.486.2336D, 2020MNRAS.491.4973D}. Several such SAMs have also been used to study the feasibility of different seeding channels as possible origins of $z\gtrsim6$ quasars. For example, \cite{2011MNRAS.416.1916V,2012MNRAS.427L..60V,2014MNRAS.444.2442V} developed the \texttt{GAMETTE-QSODUST} data constrained SAM to probe the $z\gtrsim6$ quasars and their host galaxies. This model was used in \cite{2016MNRAS.457.3356V} and \cite{2021MNRAS.506..613S}, showing that the formation of heavy seeds~($\sim10^5~M_{\odot}$) is most crucial to the assembly of the first quasars, particularly in models where the BH accretion rate is capped at the Eddington limit. \cite{2016MNRAS.458.3047P} and \cite{2017MNRAS.471..589P} showed that light seeds~($\sim10^2~M_{\odot}$) require super-Eddington accretion to grow into the $z\gtrsim6$ quasars. \cite{2021MNRAS.503.5046L} applied a semi-analytic framework on a dark matter only simulation of a $3\times10^{12}~M_{\odot}$ halo forming at $z=6$~(presumably hosting a luminous quasar); they demonstrated that the progenitors of this halo can be sites for the formation of massive DCBH seeds. 

While SAMs, being computationally inexpensive, can probe a wide range of seed models relatively quickly, they are unable to self-consistently track the hydrodynamics of gas. Cosmological hydrodynamic simulations~\citep{2012ApJ...745L..29D,2014Natur.509..177V,2015MNRAS.452..575S,2015MNRAS.450.1349K,2015MNRAS.446..521S,2016MNRAS.460.2979V,2016MNRAS.463.3948D,2017MNRAS.467.4739K,2017MNRAS.470.1121T,2019ComAC...6....2N,2020MNRAS.498.2219V} are more readily able to decipher the role of gas hydrodynamics in forming the $z\gtrsim6$ quasars~(see, e.g., the review by \citealt{2020NatRP...2...42V}). Since the $z\gtrsim6$ quasars are extremely rare, we need extremely large volumes to probe these objects~(note however that they are much more computationally expensive than SAMs). \texttt{MassiveBlack}~\citep{2012ApJ...745L..29D}, with a volume of $[0.75~\mathrm{Gpc}]^3$, revealed that $z\gtrsim6$ quasars can form in extremely massive halos~($\gtrsim10^{12}~M_{\odot}/h$ at $z\sim6$) via a steady inflow of cold gas driving sustained accretion rates close to the Eddington limit. This was further confirmed using follow-up zoom simulations at much higher resolutions~\citep{2014MNRAS.440.1865F}. \texttt{BlueTides}~\citep{2016MNRAS.455.2778F}, with a volume of $[0.5~\mathrm{Gpc}]^3$~(but higher resolution compared to \texttt{MassiveBlack}), further revealed the role of higher order features~(particularly low tidal fields, see \citealt{2017MNRAS.467.4243D}) of the initial density field in producing the fastest accretion rates necessary to assemble the $z\gtrsim7$ quasars. 

The results of \cite{2017MNRAS.467.4243D} motivated \cite{2021MNRAS.tmp.2867N}~(hereafter N21), which was a systematic study of the impact of higher order features of rare density peaks on the subsequent black hole~(BH) growth. Using the method of constrained Gaussian realizations~\citep{1991ApJ...380L...5H,1996MNRAS.281...84V}, N21 was able to generate initial conditions comprising of the rarest density peaks with the desired higher order features~(i.e. 1st and 2nd order derivatives). They demonstrated that highly compact peaks with low tidal fields led to the fastest BH growth. Due to their finite resolution however, cosmological hydrodynamic simulations are limited in terms of their ability to probe low-mass 
BH seeding channels.  Consequently, the vast majority of the simulations targeting $z\gtrsim6$ quasars described in the previous paragraph~(also including \citealt{2009MNRAS.400..100S, 2014MNRAS.439.2146C, 2016MNRAS.457L..34C, 2020arXiv201201458Z}) used simple halo based seeding prescriptions~(seeds are inserted in halos above a prescribed halo mass) that do not distinguish between different physical seeding channels. Therefore, while all these simulations have been generally successful in broadly reproducing the $z\gtrsim6$ quasars, their ability to reveal insights into the seeding environments of these objects is still limited. With upcoming LISA measurements being amongst the most promising probes for revealing the mechanism of BH seed formation, the time is ripe for developing simulations that can reliably distinguish between different BH seeding channels. 

Numerous studies have implemented 
gas-based black hole seeding prescriptions~\citep{2011ApJ...742...13B,2013MNRAS.428.2885D,2014MNRAS.442.2304H,2015MNRAS.448.1835T,2016MNRAS.460.2979V,2017MNRAS.470.1121T,2017MNRAS.467.4739K,2017MNRAS.468.3935H,2019MNRAS.486.2827D,2020arXiv200601039L,2020arXiv200204045T}. \citet{2021MNRAS.507.2012B,2021MNRAS.tmp.3110B} have recently conducted 
a systematic study  
to assess the impact of gas-based black hole seeding prescriptions on $z\gtrsim7$ SMBH populations. 
These seed models are built on the framework of the \texttt{IllustrisTNG} galaxy formation model~\citep{2017MNRAS.465.3291W,2018MNRAS.473.4077P}. They 
seeded black holes in halos via criteria based on dense, metal poor gas mass, halo mass, gas spin as well as incident Lyman Werner~(LW) flux. The resulting family of models is generally agnostic about which theoretical seeding channels they represent, but their parameters could be tuned to represent any of the seeding channels described above 
(PopIII, NSC or DCBH). 
By applying these models to zoom simulations of modestly overdense regions~($3.3\sigma$ overdensity, targeting a $\sim10^{11}~M_{\odot}/h$ halo at $z=5$), they 
found that changing different seed parameters would leave qualitatively distinct imprints on the BH merger rates. In particular, \cite{2021MNRAS.507.2012B} found that when the dense, metal poor gas mass threshold is increased, it suppresses the seeding and merger rates more strongly at $z\lesssim15$ compared to higher redshifts. On the other hand, an increase in the total halo mass threshold for seeding causes stronger suppression of seeding and merger rates at $z\sim11-25$ compared to $z\lesssim11$. These results suggest that discrepancies between the merger rates of LISA binaries will contain insights into their seeding environments. \cite{2021MNRAS.tmp.3110B} found that even when a moderately low LW flux threshold~($\gtrsim50~J_{21}$) is adopted for seeding, it can dramatically suppress seed formation and prevent the assembly of $z\gtrsim7$ SMBHs. This suggests that the bulk of the $z\gtrsim7$ SMBH population~(likely revealed by JWST and Lynx) may not originate from DCBH seeding channels.    

The zoom regions of \cite{2021MNRAS.507.2012B,2021MNRAS.tmp.3110B} were not nearly overdense enough to be possible sites for the formation of $z\gtrsim6$ quasars. In this work, we use constrained Gaussian realizations of extreme overdense regions~($\gtrsim5\sigma$ overdensities forming $\gtrsim10^{12}~M_{\odot}/h$ halos by $z\sim7$), and investigate the impact of BH seed models on the formation of the $z\gtrsim6$ quasars. Apart from the seed models, our underlying galaxy formation model is adopted from the \texttt{IllustrisTNG} simulation suite. 

Section \ref{Simulation_setup_sec} describes the simulation setup, including the main features of the \texttt{IllustrisTNG} galaxy formation model, the BH seeding and accretion models, and the generation of the constrained initial conditions. Section \ref{Results_sec} describes the main results concerning the impact of environment, seeding and accretion models on BH growth. Finally, Section \ref{Conclusions_sec} summarizes the main conclusions of our work.

\section{Simulation setup}
\label{Simulation_setup_sec}
Our simulations were run using the \texttt{AREPO} code~\citep{2010MNRAS.401..791S,2011MNRAS.418.1392P,2016MNRAS.462.2603P,2020ApJS..248...32W}, which 
includes a gravity and magneto-hydrodynamics~(MHD) solver. The simulations are cosmological in nature, which are performed within a representative portion of an expanding universe described by a fixed comoving volume~($9~\mathrm{cMpc/h}$ box size) with the following cosmology adopted from \cite{2016A&A...594A..13P}: ($\Omega_{\Lambda}=0.6911, \Omega_m=0.3089, \Omega_b=0.0486, H_0=67.74~\mathrm{km}~\mathrm{sec}^{-1}\mathrm{Mpc}^{-1},\sigma_8=0.8159,n_s=0.9667$). The code uses a PM-Tree~\citep{1986Natur.324..446B} method to solve for gravity, which is contributed by dark matter, gas, stars and BHs. Within the resulting gravitational potential, the gas dynamics is computed by the MHD solver, which uses a quasi-Lagrangian description of the fluid within an unstructured grid generated via a Voronoi tessellation of the domain.

Our galaxy formation model is adopted from the \texttt{IllustrisTNG} simulation suite~\citep{2018MNRAS.475..676S,2018MNRAS.475..648P,2018MNRAS.475..624N,2018MNRAS.477.1206N,2018MNRAS.480.5113M,2019ComAC...6....2N} \citep[see also][]{2018MNRAS.479.4056W,2018MNRAS.474.3976G,2019MNRAS.485.4817D,2019MNRAS.484.5587T,2019MNRAS.483.4140R,2019MNRAS.490.3196P,2021MNRAS.500.4597U,2021MNRAS.503.1940H}. The only substantive changes to the galaxy formation implemented here are in the sub-grid prescriptions 
for BH seeding and accretion. The remaining aspects of our galaxy formation model are the same as \texttt{IllustrisTNG} which are detailed in \cite{2017MNRAS.465.3291W} and \cite{2018MNRAS.473.4077P}; here, we provide a brief summary: 

\begin{itemize}
    \item Energy loss via radiative cooling includes contributions from primodial species~($\mathrm{H},\mathrm{H}^{+},\mathrm{He},\mathrm{He}^{+},\mathrm{He}^{++}$, based on \citealt{1996ApJS..105...19K}), as well as metals~(using pre-calculated tables for cooling rates as in \citealt{2008MNRAS.385.1443S}) in the presence of a spatially uniform, time dependent UV background. Note that cooling due to molecular Hydrogen~($H_2$) is not explicitly included in the model. 
    \item Stars are stochastically formed within gas cells with densities exceeding $0.1~\mathrm{cm}^{-3}$ with an associated time scale of $2.2~\mathrm{Gyr}$. The star forming gas cells then represent an unresolved multiphase interstellar medium, which is modeled by an effective equation of state~\citep{2003MNRAS.339..289S,2014MNRAS.444.1518V}. The model implicitly assumes that stars are produced within an unresolved cold dense component in these gas cells, which would presumably form via $H_2$ cooling.  
    \item The stellar evolution model is adopted from \cite{2013MNRAS.436.3031V} with modifications for \texttt{IllustrisTNG} as in \cite{2018MNRAS.473.4077P}. Star particles represent a single stellar population with fixed age and metallicity. The initial mass function is assumed to be \cite{2003PASP..115..763C}. The stellar evolution subsequently leads to chemical enrichment, wherein the evolution of seven species of metals~(C, N, O, Ne, Mg, Si, Fe) are individually tracked in addition to H and He.  
    
    \item Feedback from stars and Type Ia/II Supernovae are modelled as galactic scale winds~\citep{2018MNRAS.475..648P}, via which mass, momentum and metals are deposited on to the gas surrounding the star particles. 
    
\item Models for BH formation and growth are detailed in the next two subsections. The treatment of BH dynamics and mergers is the same as in  \texttt{IllustrisTNG}. Due to the limited gas mass resolution, our simulations cannot self consistently reveal the small-scale dynamics of BHs, particularly at their lowest masses. To stabilize the BH dynamics, they are ``re-positioned" to the nearest potential minimum within its ``neighborhood"~(defined by $10^3$ nearest neighboring gas cells). As a result, a BH is also promptly merged when it is within the neighborhood of another BH.     
\end{itemize}
\subsection{Black hole seeding}
\label{Black hole seeding}
We consider a range of BH seeding prescriptions, which include the default halo based seeding prescription of \texttt{IllustrisTNG} where seeds of mass $8\times10^5~M_{\odot}$ are inserted in halos which exceed a threshold mass of $5\times10^{10}~M_{\odot}/h$ and do not already contain a BH~(hereafter referred to as the ``TNG seed model"). 

Additionally, we explore the gas-based seeding prescriptions developed in \cite{2021MNRAS.507.2012B} and \cite{2021MNRAS.tmp.3110B}. These are comprised of a combination of seeding criteria based on various gas properties of halos. These criteria are designed such that our overall family of seed models broadly encompasses popular theoretical channels such as Pop III, NSC and DCBH seeds, all of which exclusively form in regions comprised of dense, metal poor gas.  Here we briefly summarize them as follows: 
\begin{itemize}
\item \textit{Dense, metal poor gas mass criterion:} Seeds can only form in halos that exceed a threshold for dense~($>0.1~\mathrm{cm}^{-3}$), metal poor~($Z<10^{-4}~Z_{\odot}$) gas mass, specified by $\tilde{M}_{\mathrm{sf,mp}}$ in the units of the seed mass $M_{\mathrm{seed}}$.
\item \textit{Halo mass criterion:} Seeds can only form in halos that have exceeded a threshold for the total halo mass, specified by $\tilde{M}_{h}$ in the units of the seed mass $M_{\mathrm{seed}}$.

\item \textit{LW flux criterion:} In selected models, we also require the dense, metal poor gas to be exposed to Lyman Werner~(LW) fluxes above a critical value $J_{\mathrm{crit}}$. More specifically, seeds only form in halos with a minimum threshold for dense, metal poor, LW illuminated gas mass, denoted by  $\tilde{M}_{\mathrm{sf,mp,LW}}$ in the units of the seed mass $M_{\mathrm{seed}}$. Star formation is suppressed in these seed forming regions. Given that our simulations do not contain full radiative transfer, the LW flux from Pop III and Pop II stars is computed using an analytic prescription described in \cite{2021MNRAS.tmp.3110B}.
\end{itemize}

Our seed model is therefore described by four parameters, namely $\tilde{M}_{\mathrm{sf,mp}}$, $\tilde{M}_{\mathrm{h}}$, $J_{\mathrm{crit}}$ and $M_{\mathrm{seed}}$. All of our simulations include the first two parameters, and throughout the text 
the \textit{dense, metal poor gas mass criterion} and \textit{halo mass criterion} are labelled as \texttt{SM*_FOF*} where the `*'s correspond to the values of $\tilde{M}_{\mathrm{sf,mp}}$ and $\tilde{M}_{\mathrm{h}}$. For example, $\tilde{M}_{\mathrm{sf,mp}}=5$ and $\tilde{M}_{\mathrm{h}}=3000$ will correspond to \texttt{SM5_FOF3000}. Runs which additionally apply the \textit{LW flux criterion} contain an extra suffix \texttt{LW*} where `*' corresponds to $J_{\mathrm{crit}}$; for example, if a criterion with $J_{\mathrm{crit}}=300~J_{21}$ is added to \texttt{SM5_FOF3000}, it will be labeled as \texttt{SM5_FOF3000_LW300}. Lastly, the seed mass $M_{\mathrm{seed}}$ will be explicitly stated in the text and figure legends.

\subsection{BH accretion and feedback models}
Black holes grow via a modified Bondi-Hoyle accretion prescription, with 
the maximum accretion rate limited to some factor $f_{\rm Edd} \geq 1$ times 
the Eddington accretion rate (which we refer to as the `Eddington factor'):
\begin{eqnarray}
\dot{M}_{\mathrm{BH}}=\mathrm{min}(\alpha \dot{M}_{\mathrm{Bondi}}, \: f_{\mathrm{edd}}\dot{M}_{\mathrm{Edd}})\\
\dot{M}_{\mathrm{Bondi}}=\frac{4 \pi G^2 M_{\mathrm{BH}}^2 \rho}{c_s^3}\\
\dot{M}_{\mathrm{Edd}}=\frac{4\pi G M_{\mathrm{BH}} m_p}{\epsilon_r \sigma_T~c}
\label{bondi_eqn}
\end{eqnarray} 
$\alpha$ is referred to as the `Bondi boost' factor which is often used to boost the accretion rate to account for the inability to resolve the small scale vicinity of the BH. $G$ is the gravitational constant, $\rho$ is the local gas density, $M_{\mathrm{BH}}$ is the BH mass, $c_s$ is the local sound speed, $m_p$ is the proton mass, and $\sigma_T$ is the Thompson scattering cross section.  In practice, ``local'' quantities are calculated as the kernel-weighted averages over nearby particles, typically those within a few $\times h^{-1} \mathrm{pc}$.  Accreting black holes radiate at luminosities given by, 
\begin{equation}
    L=\epsilon_r \dot{M}_{\mathrm{BH}} c^2,
    \label{bol_lum_eqn}
\end{equation}
where $\epsilon_r$ is the radiative efficiency.

In the IllustrisTNG implementation, AGN feedback occurs both in `thermal mode' as well as `kinetic mode'. For Eddington ratios~($\eta \equiv \dot{M}_{\mathrm{bh}}/\dot{M}_{\mathrm{edd}}$) higher than a critical value of  $\eta_{\mathrm{crit}}=\mathrm{min}[0.002(M_{\mathrm{BH}}/10^8 M_{\odot})^2,0.1]$, thermal energy is deposited on to the neighboring gas at a rate of $\epsilon_{f,\mathrm{high}} \epsilon_r \dot{M}_{\mathrm{BH}}c^2$ where $ ~\epsilon_{f,\mathrm{high}} \epsilon_r=0.02$. $\epsilon_{f,\mathrm{high}}$ is called the ``high accretion state" coupling efficiency. If the Eddington ratio is lower than the critical value, kinetic energy is injected into the gas at irregular time intervals, which manifests as a `wind' oriented along a randomly chosen direction. The injected rate is $\epsilon_{f,\mathrm{low}}\dot{M}_{\mathrm{BH}}c^2$ where $\epsilon_{f,\mathrm{low}}$ is called the `low accretion state' coupling efficiency~($\epsilon_{f,\mathrm{low}} \lesssim 0.2$). For further details, we direct the interested readers to \cite{2017MNRAS.465.3291W}. 

The main parameters of our accretion model include the Bondi boost $\alpha$, the radiative efficiency $\epsilon_r$ and the Eddington factor $f_{\mathrm{edd}}$. The default values adopted in the \texttt{IllustrisTNG} suite are $\alpha=1,\epsilon_r=0.2~\&~f_{\mathrm{edd}}=1$. We largely use this accretion model, and hereafter refer to it as the ``TNG accretion model''. However, we also run some simulations with different variations of these parameters, particularly when comparing our results to other studies. These variations include different combinations of $\alpha=1~\&~100$, $\epsilon_r=0.2~\&~0.1$ and  $f_{\mathrm{edd}}=1-100$. In the figure legends, these are labelled as \texttt{Boost*_RadEff*_EddFac*} where the `*'s correspond to values of $\alpha$, $\epsilon_r$ and $f_{\mathrm{edd}}$ respectively.

\subsection{Initial Conditions: constrained Gaussian realizations}

We expect the brightest $z>6$ quasars to live in the rarest and most extreme overdensities in the Universe. In order to create initial conditions~(ICs) that can produce such regions within a relatively small $9~\mathrm{cMpc}/h$ box, we apply the technique of \textit{constrained Gaussian realizations} (CR). 
The CR method can efficiently sample a Gaussian Random field conditioned on various (user-specified) large-scale features.
This technique was originally introduced by \cite{1991ApJ...380L...5H} and \cite{1996MNRAS.281...84V}. We use the most recent implementation of this technique, i.e. the \href{https://github.com/yueyingn/GaussianCR}{\texttt{GaussianCR}} code to generate the initial conditions. This code was fully developed by N21, wherein it was extensively tested against large volume uniform cosmological  simulations in terms of reproducing the halo assembly, star formation and BH growth histories. Here we briefly summarize the main features for completeness, while the full details of the underlying formalism are described in N21.   

Overall, \href{https://github.com/yueyingn/GaussianCR}{\texttt{GaussianCR}} constrains 18 parameters at the peak location~(see N21 for details). In this work, we vary three parameters that were shown by N21 to be most consequential to BH growth. These are the following:
\begin{itemize}
    \item The peak height `$\nu$' quantifies the `rarity' of the peak by specifying its height in the units of the variance of $\delta_G (\textbf{r})$ denoted by $\sigma_{R_G}$, i.e.
    \begin{eqnarray}
        \nu \equiv \delta(\textbf{r}_{\mathrm{peak}})/\sigma_{R_G}, \\
        \sigma_{R_G}^2\equiv\left<\delta_G (\textbf{r}) \delta_G (\textbf{r})\right>= \int \frac{P(k)}{(2\pi)^3} \hat{W}^2(k,R_G) dk 
    \end{eqnarray}
    where $\textbf{r}_{\mathrm{peak}}$ is the peak position. $P(k)$ is the power spectrum and $\tilde{\delta}_G(\textbf{r})\equiv \int \tilde{\delta} (\textbf{r}) W(\textbf{r},R_G)$ is the overdensity smoothened over a scale $R_G$ using the Gaussian window function $W(\textbf{r},R_G)= \exp(-r^2/2R_G^2)$ and its Fourier transform $\hat{W}(k,R_G)=\exp(-k^2 R_G^2/2)$. 

    \item The peak compactness `$x_d$' is set by the second order derivatives of the smoothed overdensity field. More quantitatively, it is determined by the eigenvalues of the Hessian matrix $\partial_{ij}\delta_{G}$~($i,j=1,2~\&~3$ are $x,y~\&~z$ components respectively) can be parametrized as 
    \begin{eqnarray}
        \lambda_1= \frac{x_d \sigma_2(R_g)}{1+a_{12}^2+a_{13}^2}\\
        \lambda_2= a_{12}^2 \lambda_1\\
        \lambda_3= a_{13}^2 \lambda_1
    \end{eqnarray}
    where $\sigma_2(R_g) \equiv \int \frac{P(k)}{(2\pi)^3} \hat{W}^2(k,R_G) k^2 dk$, and $a_{12}$ and $a_{13}$ are the axis ratios that determine the shape of the mass distribution around the ellipsoidal peak.

    \item Lastly, the tidal strength $\epsilon$ is determined by second order derivative of the gravitational potential i.e. the tidal tensor $T_{ij}~(i=1,2,3)$. The eigenvalues of the tidal tensor can be parametrized by, 
    \begin{equation}
        \left[\epsilon \cos{\frac{\omega+2\pi}{3}}, \epsilon \cos{\frac{\omega-2\pi}{3}}, \epsilon \cos{\frac{\omega}{3}} \right],
    \end{equation} 
    where $\epsilon$ determines the overall magnitude of the tidal tensor, and $\omega$ determines the relative strengths of the tidal tensor along the three eigenvectors.
\end{itemize}

\subsubsection{Our choice of peak parameters}
\begin{table*}
\centering
\begin{tabular}{|c|c|c|c|c|}
      IC & $R_G~(\mathrm{Mpc}/h)$ & $\nu~(\sigma_{R_G})$ & $x_d~(\sigma_2)$ & $\epsilon~(\mathrm{km~s^{-1}~\mathrm{Mpc}^{-1}})$ \\
      \hline
     \texttt{5SIGMA} & 1.0 & 5 & 3.6~(ave) & 34.0~(ave)  \\
     \texttt{5SIGMA_COMPACT} & 1.0 & 5 & 5.8~($+3\sigma$) & 15.0~($-2\sigma$)\\
     \texttt{6SIGMA} & 1.3 & 6 & 4.0~(ave) & 34.0~(ave)\\
     \hline
\end{tabular}
\caption{The adopted values for the peak parameters for $\texttt{5SIGMA}$, $\texttt{5SIGMA_COMPACT}$ and $\texttt{6SIGMA}$. These constrained initial conditions~(ICs) are characterized by the  smoothing scale $R_G$~(2nd col), the peak height $\nu$~(3rd col), the peak compactness $x_d$~(4th col) and the tidal scalar $\epsilon$~(5th col); these 4 parameters are most consequential to BH growth. The remaining parameters do not significantly impact BH growth, and have been fixed to be the typical values of their underlying distributions~(see Figure 5 of N21). }
\label{peak_parameters_table}
\end{table*}
N21 shows that BH growth is the most efficient within rare peaks~(high $\nu$) that are compact~(high $x_d$) and allow for gas infall to occur from all directions~(low tidal strength $\epsilon$).
  Therefore, throughout this paper, we make intentional choices for $\nu$, $x_d$ and $\epsilon$ and fix the remaining parameters at their most probable values~(see Figure 5 of N21) . Table \ref{peak_parameters_table} summarizes the adopted  parameter values for $\nu$, $x_d$ and $\epsilon$. More specifically, we look at the following three regions:
\begin{itemize}
    \item  We choose a $5\sigma$ peak~($\nu=5$) at scales of $R_G=1~\mathrm{Mpc}/h$, with $x_d$ and $\epsilon$ corresponding to the typical values i.e. the maxima of their respective distributions. The peak height was chosen to produce a target halo mass of $10^{12}~M_{\odot}/h$ at $z=7$. It is hereafter referred to as \texttt{5SIGMA}.
    \item We again choose a $5\sigma$ peak at $R_G=1~\mathrm{Mpc}/h$, but with a compactness $x_d$ that is $3\sigma$ away from the typical value, and a tidal strength $\epsilon$ that is $-2\sigma$ away from the mean value. This also targets the assembly of a $10^{12}~M_{\odot}/h$ halo at $z=7$, and is referred to as \texttt{5SIGMA_COMPACT}. 
    \item Lastly, we choose a $6\sigma$ peak~($\nu=6$) at scales of $R_G=1.3~\mathrm{Mpc}/h$ with typical values for $x_d$ and $\epsilon$. This targets a $5\times10^{12}~M_{\odot}/h$ halo at $z=7$, and is referred to as \texttt{6SIGMA}.
\end{itemize}
Note that the target halos produced in \texttt{6SIGMA} and \texttt{5SIGMA_COMPACT} regions have number densities roughly similar to those of the observed $z\sim6$ quasars~($\sim1~\mathrm{Gpc}^{-3}$). 
In contrast, \texttt{5SIGMA} produces a target halo that is $\sim100$ times more common.

Finally, we also note that the BH growth can depend on the specific realization of the large scale density field. However, upon exploring 5 different realizations for a select few BH models, we found that the differences in BH growth were mild~($z\sim6$ BH masses vary by factors $\lesssim2$). 


\subsection{Simulation resolution}

In \cite{2021MNRAS.507.2012B,2021MNRAS.tmp.3110B}, we performed detailed resolution convergence tests for our BH seed model and found that for our fiducial model with $\tilde{M}_{\mathrm{sf,mp}}=5$ and $\tilde{M}_{h}=3000$, the seed formation rates are reasonably well converged for gas mass resolutions $\lesssim10^4~M_{\odot}/h$. The resolution convergence becomes slower as the models are made more restrictive by increasing $\tilde{M}_{\mathrm{sf,mp}}$ or by introducing a LW flux criterion. As we shall see in Section \ref{BH growth at higher resolutions_sec}, the resolution convergence properties of our constrained regions are similar to that of the zoom region of \cite{2021MNRAS.507.2012B,2021MNRAS.tmp.3110B}.

To achieve a gas mass resolution of $\sim10^4~M_{\odot}/h$ in a box size of $9~\mathrm{Mpc}/h$, we need $N=720$ DM particles per dimension~(note that the number of gas cells are initially assigned to be equal to the DM particles, but as the simulations evolve the gas cells can undergo refinement or de-refinement). However, running such a simulation until $z=6$ requires a substantial amount of computing time and memory, particularly in regions with extreme overdensities. Therefore, to facilitate a rapid exploration of the large parameter space of our seed models, we choose $N=360$. We assign this to be our fiducial resolution, and it corresponds to a gas mass resolution of $\sim10^5~M_{\odot}/h$. Note that this is only slightly lower than the highest resolution box of the \texttt{IllustrisTNG} suite i.e. TNG50. That being said, we do run higher resolution realizations~($N=720$) for a few selected models, particularly those that successfully produce a $z\gtrsim6$ quasar. As we shall see in Section \ref{BH growth at higher resolutions_sec}, the final BH mass at $z\lesssim7$ is not significantly impacted by resolution. Additionally, we use the $N=720$ runs to probe the lowest seed mass considered in this work i.e. $M_{\mathrm{seed}}=1.25\times10^4~M_{\odot}/h$. The fiducial $N=360$ run can only probe seed masses of $M_{\mathrm{seed}}=1\times10^5~\&~8\times10^5~M_{\odot}/h$. Hereafter, unless otherwise stated, we are using $N=360$. For runs that use $N=720$, it shall be explicitly stated in the captions or the text.

\section{Results}

\label{Results_sec}

\subsection{Halo environments: evolution from $z\sim20-6$}

\begin{figure*}
\includegraphics[width=5.2 cm]{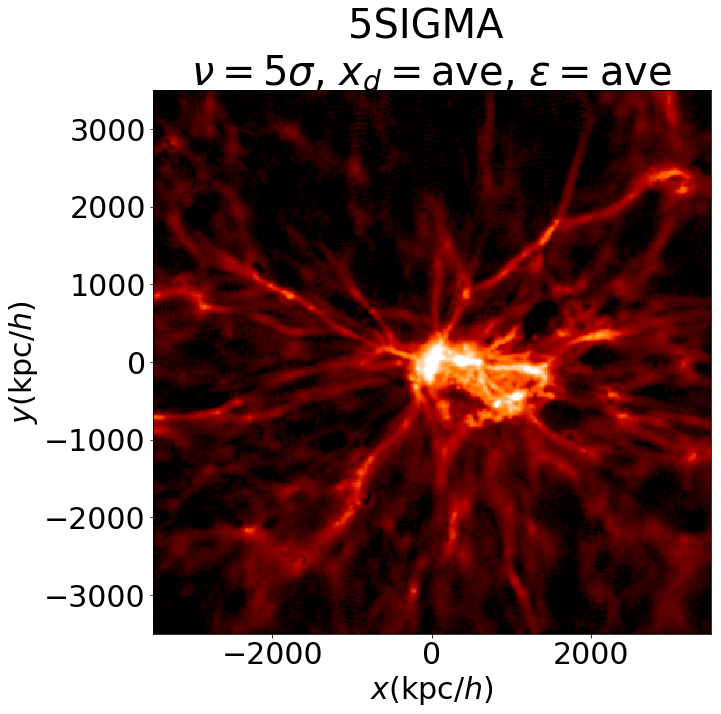}
\includegraphics[width=5.2 cm]{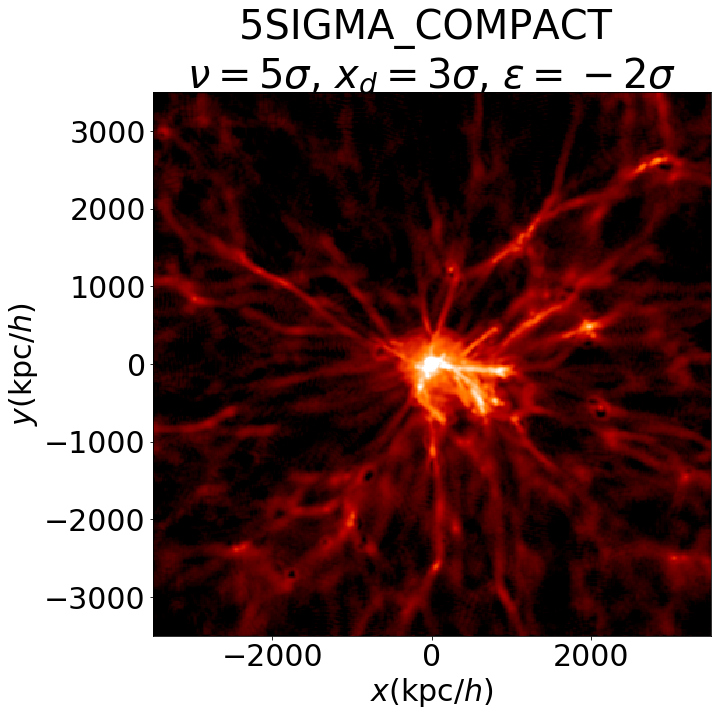}
\includegraphics[width=6.5 cm]{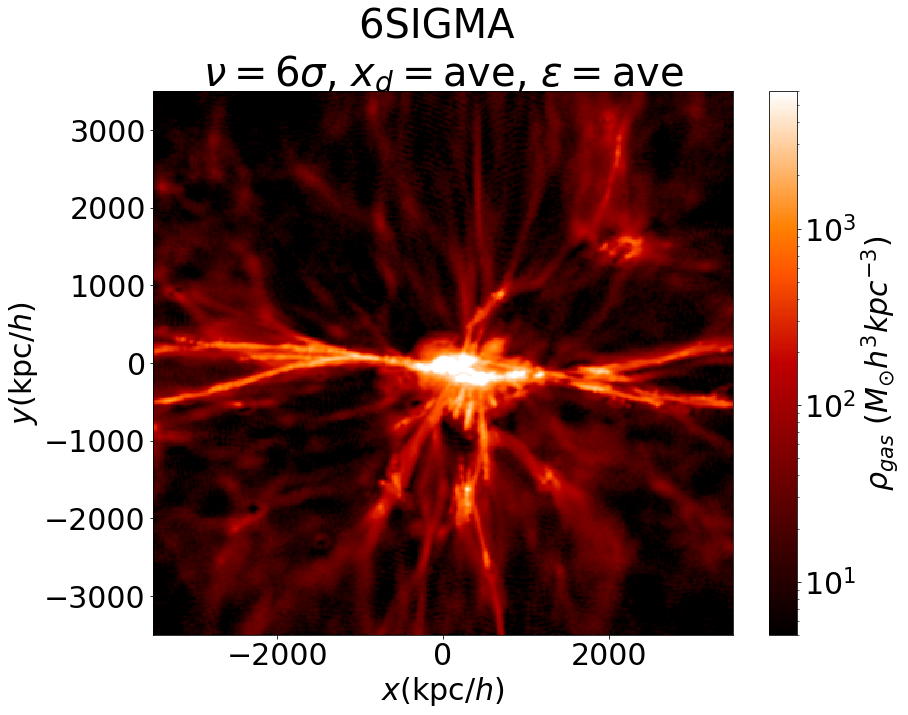}
\caption{Gas density field at the $z=6$ snapshot for the three constrained Gaussian initial conditions we explored in this work. The left panel is centered at a $5\sigma$ overdensity peak at $R_{G}=1~\mathrm{Mpc}/h$, and typical values for compactness $x_d$ and tidal strength $\epsilon$; this is hereafter referred to as \texttt{5SIGMA}. The middle panel is centered at a $5\sigma$ overdensity peak with $+3\sigma$ higher compactness, and $-2\sigma$ lower tidal strength; we refer to this as \texttt{5SIGMA_COMPACT}. The right panel is centered at a $6\sigma$ overdensity peak at $R_{G}=1.3~\mathrm{Mpc}/h$ with average values of compactness and tidal strength; this is referred to as \texttt{6SIGMA}.  We can see that the gas distribution in \texttt{5SIGMA_COMPACT} is more isotropic and centrally concentrated compared to \texttt{5SIGMA} and \texttt{6SIGMA}.}
\label{halo_properties_fig}
\end{figure*}

\begin{figure}
\includegraphics[width=8 cm]{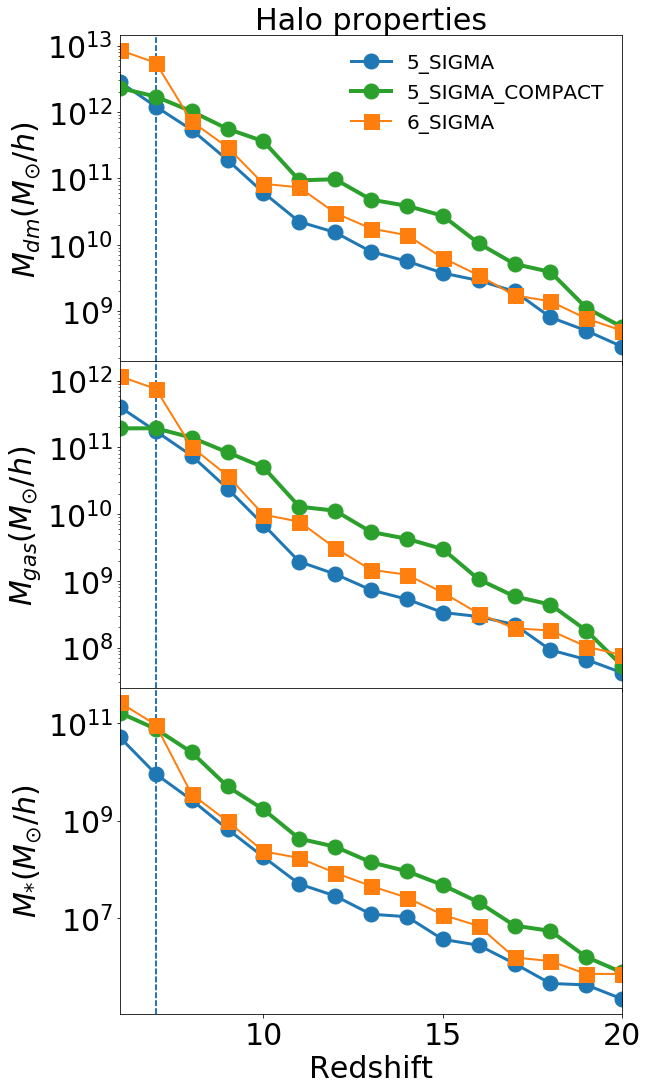}
\caption{The evolution of the most massive halo~(MMH) from $z\sim20$ to $z\sim7$ for \texttt{5SIGMA}~(blue), \texttt{5SIGMA_COMPACT}~(green) and \texttt{6SIGMA}~(orange) lines. The top, middle and bottom panels show the total halo mass, gas mass and stellar mass respectively. For \texttt{5SIGMA} and \texttt{5SIGMA_COMPACT}, we reach our target halo mass of $\sim10^{12}~M_{\odot}/h$ at $z=7$; they both assemble a total gas mass of $2\times10^{11}~M_{\odot}/h$. Likewise, the \texttt{6SIGMA} volume assembles the desired target halo mass of $6\times10^{12}~M_{\odot}/h$ at $z=7$, and a gas mass of $5\times10^{11}~M_{\odot}/h$. \texttt{5SIGMA_COMPACT} assembles a stellar mass of $\sim10^{11}~M_{\odot}/h$ at $z=7$, similar to that of \texttt{6SIGMA}; this is significantly higher than \texttt{5SIGMA} which assembles a stellar mass of $9\times10^{9}~M_{\odot}/h$. More compact peaks~(at fixed peak height) lead to enhanced star formation.}
\label{halo_evolution_fig}
\end{figure}

\begin{figure}
\includegraphics[width=8.5 cm]{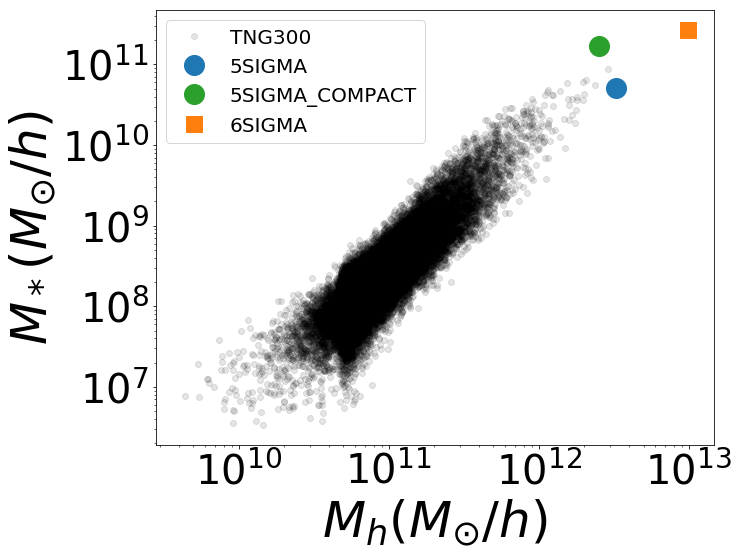}
\caption{Stellar mass vs halo mass relation at $z=6$ for the MMHs of \texttt{5SIGMA_COMPACT}~(green), \texttt{5SIGMA}~(blue) and \texttt{6SIGMA}~(orange) respectively, compared with the full halo population of TNG300~(a $[300~\mathrm{Mpc}]^3$ box---the largest volume in the \texttt{IllustrisTNG} simulation suite; faded grey circles). The predictions from the constrained runs are broadly consistent with the trends extrapolated from the TNG300 results. The \texttt{5SIGMA} and \texttt{6SIGMA} runs produce stellar mass predictions close to the mean trend, whereas the \texttt{5SIGMA_COMPACT} run produces a somewhat overly massive galaxy~(stellar mass) compared to its host halo.}
\label{SM_HM_relation_fig}
\end{figure}

\begin{figure*}
\includegraphics[width=18 cm]{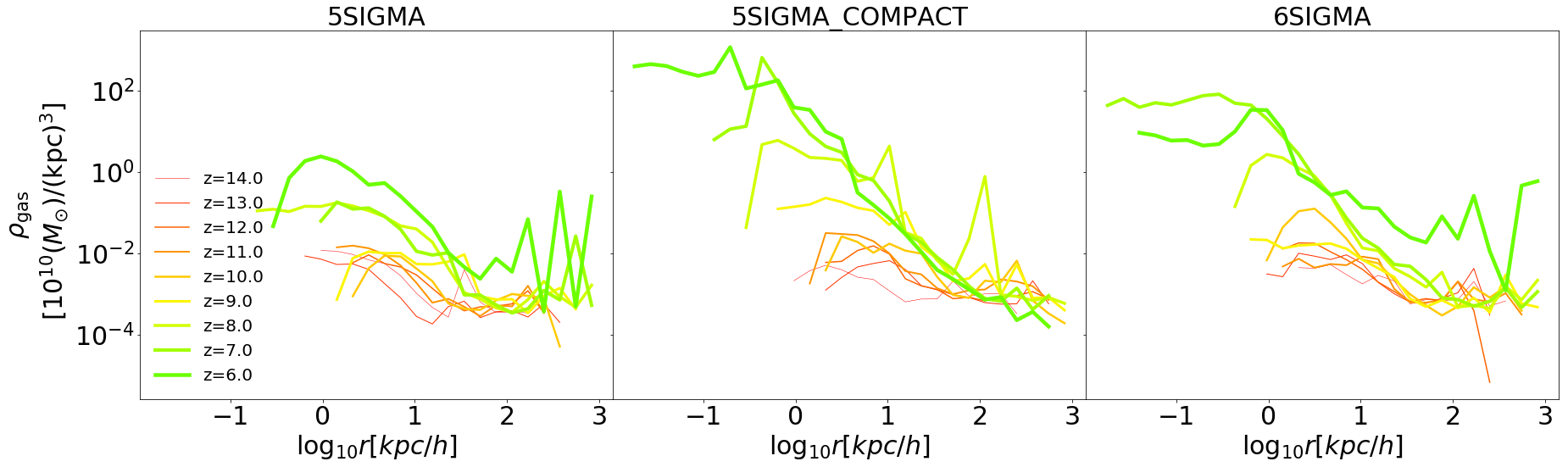}
\caption{Radially averaged 1D profiles of the gas density~(physical units), centered around the most massive BH in the MMH of our simulation volume for \texttt{5SIGMA}, \texttt{5SIGMA_COMPACT} and \texttt{6SIGMA}. Lines of different colors show the redshift evolution from $z=14$ to $z=6$. We see that gas density in the central regions steeply increases with time between $z\sim9-6$. The steepest increase is seen for \texttt{5SIGMA_COMPACT}.}
\label{1D_density_distributions_fig}
\includegraphics[width=19 cm]{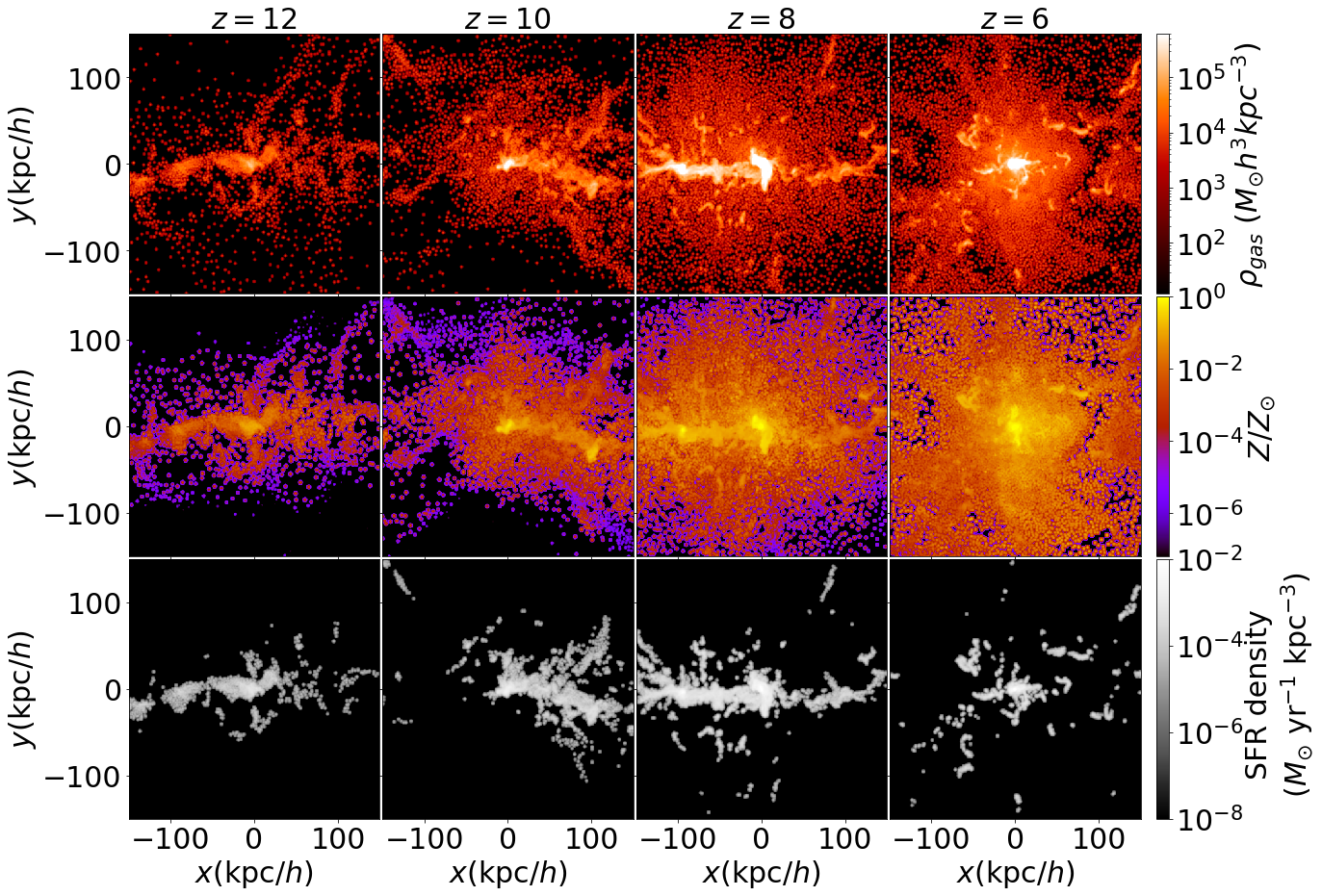}

\caption{2D profiles of the gas density, metallicity and the star formation rate density within the vicinity of the most massive BH in the MMH of the \texttt{5SIGMA_COMPACT} run. We compute these quantities averaged over a slice of thickness $10~\mathrm{kpc}/h$. Left to right panels show the redshift evolution from $z=10$ to $z=6$. At $z\lesssim9$, the steep increase in the gas density leads to increase in the star formation and metal enrichment in the halo.} 
\label{FullMapAllFieldsconstrained_fig}
\end{figure*}

Before looking at the properties of BHs, we first look at the environments in which they form and grow. Figure \ref{halo_properties_fig} shows the $z=6$ gas density profiles centered at the location of the constrained  \texttt{5SIGMA}, \texttt{5SIGMA_COMPACT} and \texttt{6SIGMA} peaks. Visually, we can clearly see that the gas distribution around the  \texttt{5SIGMA_COMPACT} peak is more compact and isotropic compared to that of \texttt{5SIGMA} and \texttt{6SIGMA}. Figure \ref{halo_evolution_fig} shows the evolution of the most massive halo~(MMH) from $z\sim20-6$, in terms of the total mass, gas mass and stellar mass. We see that both \texttt{5SIGMA_COMPACT} and \texttt{5SIGMA} runs assemble their target mass of $\sim10^{12}~M_{\odot}/h$ by $z=7$, which grows to become $\sim2\times10^{12}~M_{\odot}/h$ by $z=6$. The \texttt{6SIGMA} run assembles halo masses of $\sim7\times10^{12}~M_{\odot}/h$ and $\sim10^{13}~M_{\odot}/h$ by $z=6$.

Interestingly, the halo assembly history~(see Figure \ref{halo_evolution_fig}: top panel) of the three regions shows that for \texttt{5SIGMA_COMPACT}, the MMHs at $z\gtrsim10$ are $\sim5-10$ times more massive compared to that of \texttt{5SIGMA}~(as well as \texttt{6SIGMA}). But at $z\lesssim10$, the halo growth rate for the \texttt{5SIGMA_COMPACT} run becomes slower compared to the \texttt{5SIGMA} run, thereby explaining the similar final halo masses that both the runs assemble at $z\sim6-7$. That is likely because the MMH in the \texttt{5SIGMA_COMPACT} peak becomes more isolated at $z\lesssim10$~(after having merged with most of its neighboring massive halos by $z\sim10$). However, the \texttt{5SIGMA_COMPACT} MMH continues to become more dense during $z\sim9-6$~(due to continued gravitational collapse), which likely causes the higher compactness of \texttt{5SIGMA_COMPACT} peak compared to \texttt{5SIGMA}~(as well as \texttt{6SIGMA}) at $z\sim6-7$.

The evolution of the gas mass~(see Figure \ref{halo_evolution_fig}: middle panel) mirrors that of the total halo mass at $z\sim20-10$. More specifically, the gas mass of the MMH in \texttt{5SIGMA_COMPACT} is $\sim5-10$ times higher than that of \texttt{5SIGMA} and \texttt{6SIGMA}~(similar to the total halo mass) at $z\sim20-10$. Notably, we find that at $z\sim9-6$, there is no significant increase in the gas mass for the MMH in  \texttt{5SIGMA_COMPACT}, unlike the MMHs of \texttt{5SIGMA} and \texttt{6SIGMA}. As a result, by $z=6$, the MMH in \texttt{5SIGMA_COMPACT} ends up with a lower gas mass compared to \texttt{5SIGMA} and \texttt{6SIGMA}. As we shall see, this is happening because the gas in \texttt{5SIGMA_COMPACT} is being rapidly consumed by star formation and BH accretion, more so than \texttt{5SIGMA} and \texttt{6SIGMA}. 
The enhanced star formation in \texttt{5SIGMA_COMPACT} can be seen in the bottom panel of Figure \ref{halo_evolution_fig}, wherein the stellar mass is $\sim7$ times higher than that of \texttt{5SIGMA} at $z\sim6-7$~(making it similar to the stellar mass produced by \texttt{6SIGMA} at $z\sim6-7$).  
The \texttt{5SIGMA_COMPACT} region therefore produces an overly massive 
galaxy for its host halo mass, as clearly seen in Figure \ref{SM_HM_relation_fig}. Figure \ref{SM_HM_relation_fig} also demonstrates that our constrained runs are consistent with the stellar mass vs. halo mass relations predicted by TNG300, thereby validating this technique.

Having discussed the evolution of the global properties of the MMH, we now focus on the evolution of their internal gas distributions, which are much more consequential to BH growth. The evolution of the radially averaged gas density profiles from $z\sim14-6$ is shown in Figure \ref{1D_density_distributions_fig}. In all 
three regions, 
little evolution occurs at $z\gtrsim9$, and the central $\sim 1$ kpc/$h$ is unresolved owing to low gas densities. Between $z\sim9-6$ however, the gas densities start to significantly increase, particularly close to the halo centers. Specifically, while the overall gas mass of the MMH only increases by factors of $\sim50$ between $z\sim9-6$~(Figure \ref{halo_evolution_fig}: middle panel), the central gas densities increase by factors of $\sim100-10000$ during the same time interval. Amongst the three regions, \texttt{5SIGMA_COMPACT} shows the steepest increase in density between $z\sim9-6$. Figure \ref{FullMapAllFieldsconstrained_fig} shows the 2D color maps of the evolution of the gas density, star formation rates and metallicity for \texttt{5SIGMA_COMPACT} region at redshift snapshots of $z=12,10,8~\&~6$. We can clearly see that the steep increase in central gas density leads to a commensurate boost in the star formation, as well as metal enrichment in the central regions of the halo. 
As we shall see, this increase in the central gas densities leads to substantially increased importance of accretion-driven BH growth at $z\sim9-6$.

\subsection{BH growth in different halo environments: Impact of BH seeding models}
\begin{figure*}
\includegraphics[width=16 cm]{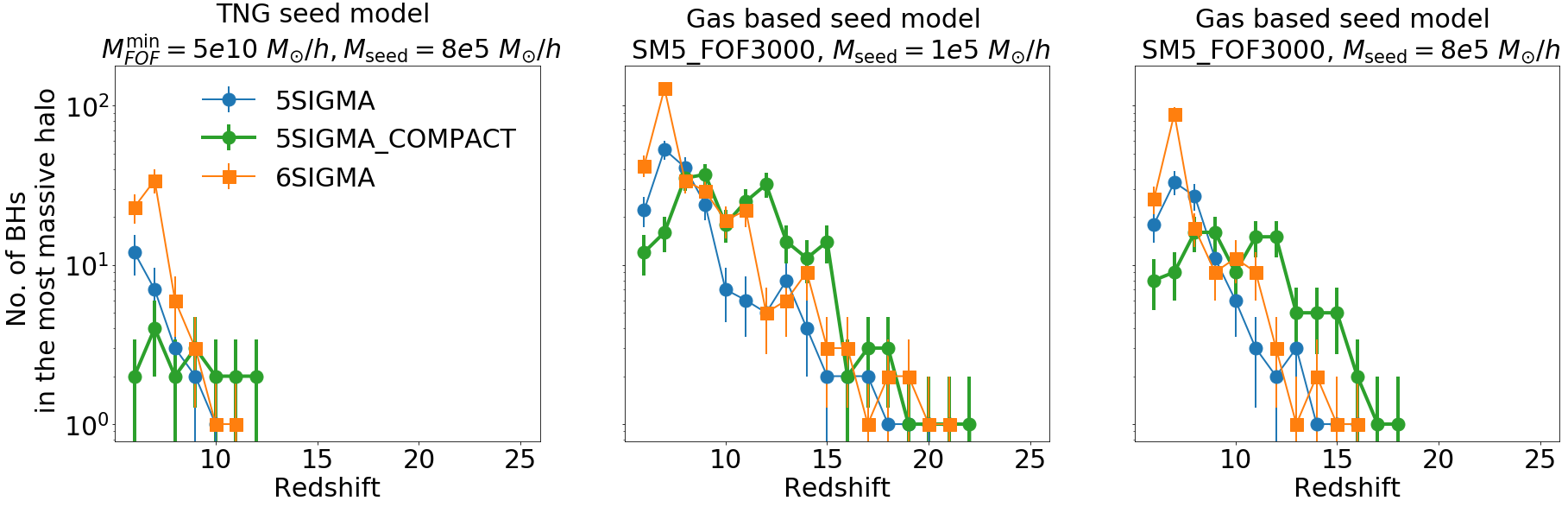}
\caption{Number of BHs present in the most massive halo~(MMH) at different redshift snapshots for \texttt{5SIGMA}~(blue), \texttt{5SIGMA_COMPACT}~(green) and \texttt{6SIGMA}~(orange) lines. Error-bars correspond to Poisson errors. In the leftmost panel, we use the default seeding prescription from the IllustrisTNG simulation suite~(referred to as TNG seed model). In the middle and right panels, we use the gas-based seeding prescription with $\tilde{M}_h=3000$ and $\tilde{M}_{\mathrm{sf,mp}}=5$~(\texttt{SM5_FOF3000}). All these runs use the default accretion prescription from the IllustrisTNG simulation suite~(TNG accretion model). We see that the onset of seed formation happens earliest within the \texttt{5SIGMA_COMPACT} run; therefore, it contains the highest number of BHs around $z\gtrsim10$. However, between $z\sim9-6$, the \texttt{5SIGMA_COMPACT} does not acquire many new BHs. On the other hand, the \texttt{5SIGMA} and \texttt{6SIGMA} MMHs continue to acquire new BHs from nearby halos between $z\sim9-6$. Therefore, by $z=6$, the \texttt{5SIGMA_COMPACT} peak has the least number of BHs and \texttt{6SIGMA} peak has the highest number of BHs.}

\label{No_of_BHs_fig}

\includegraphics[width=5.7 cm]{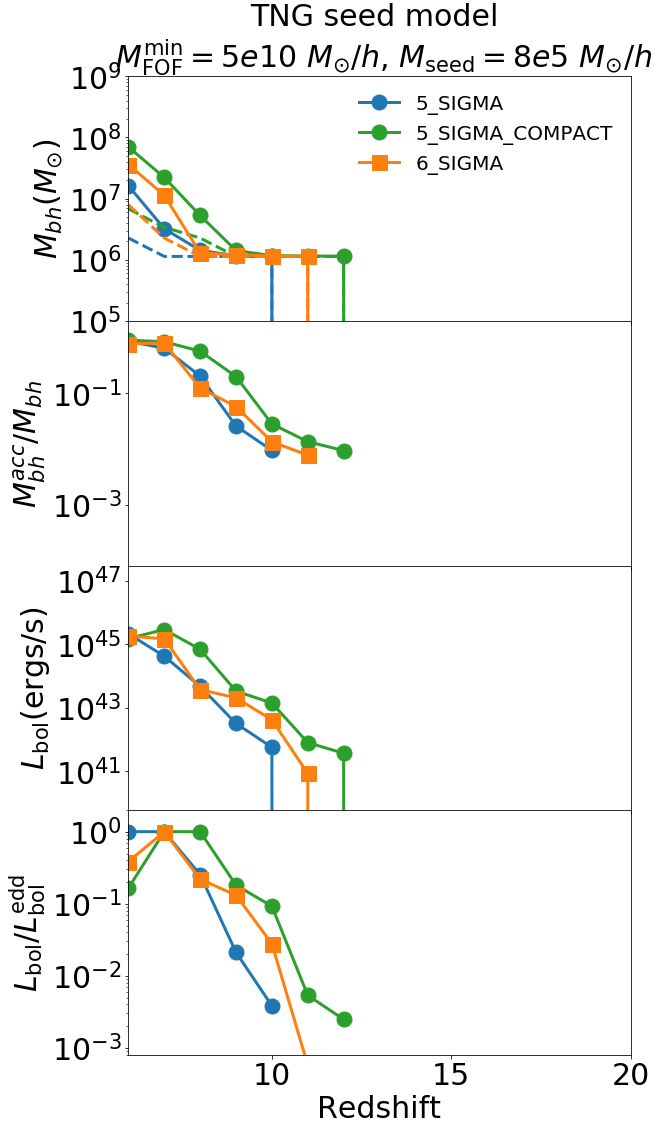}
\includegraphics[width=5.7 cm]{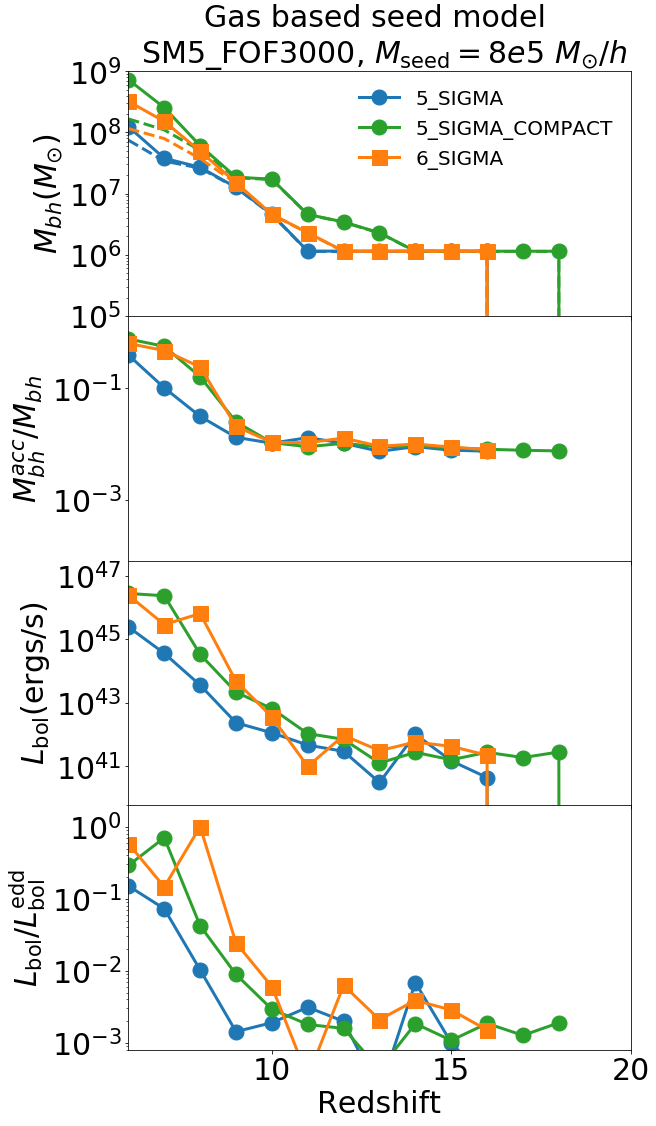}
\includegraphics[width=5.7 cm]{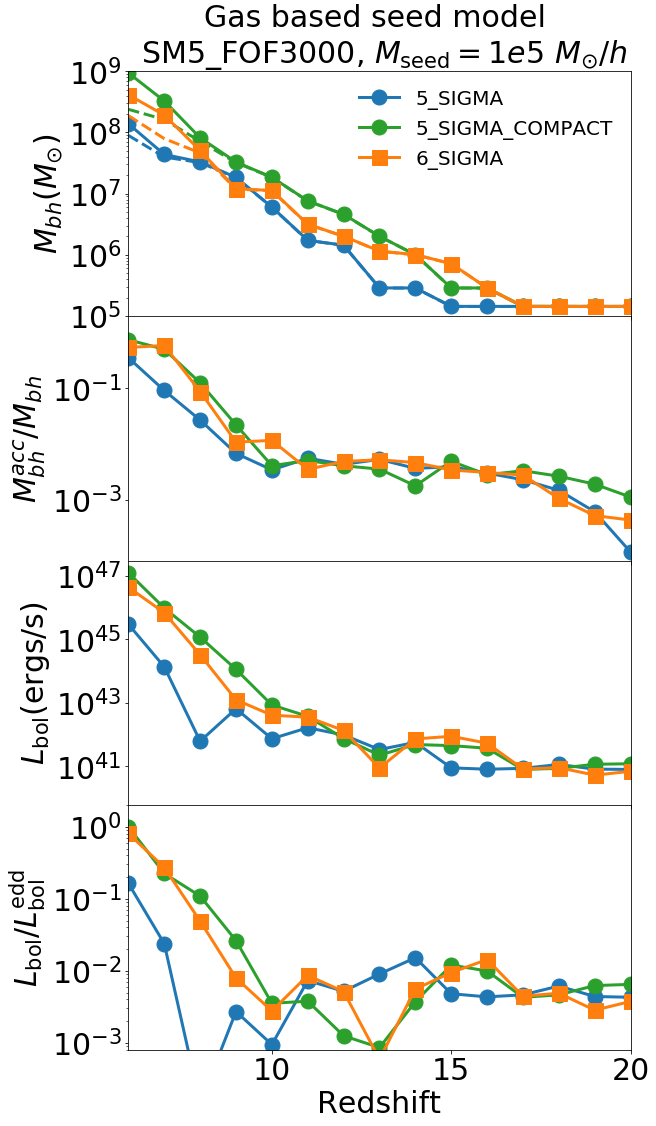}
\caption{Evolution of the most massive BH at $z=6$ in the MMH of 
\texttt{5SIGMA}~(blue), \texttt{5SIGMA_COMPACT}~(green) and \texttt{6SIGMA}~(orange) lines (hereafter, all growth histories plotted for different runs are for this particular BH in each simulation). The 1st row corresponds to the BH mass; solid lines show the total BH mass and the dashed lines show the mass accumulated only by mergers. The 2nd row shows the fraction of the current mass accumulated by gas accretion. The 3rd and 4th rows show the total bolometric luminosity in units of $\mathrm{ergs~s^{-1}}$ and the Eddington luminosity, respectively. All the runs use the TNG accretion model. In the left panels, we use the TNG seed model. In the middle and right panels, we use the gas-based seeding prescription \texttt{SM5_FOF3000}. The BH seed mass is $8 \times 10^5 M_{\odot}/h$ and $1 \times 10^5 M_{\odot}/h$ in the middle and right panels, respectively. Among the constrained volumes we explore, \texttt{5SIGMA_COMPACT} assembles the highest BH mass in all cases. Even in this region, the TNG seed model achieves a maximum BH mass of $\sim7\times10^7~M_{\odot}/h$ by $z=6$, which is significantly smaller than the typical masses of observed $z\sim6$ quasars. In contrast, the \texttt{SM5_FOF3000} models are able to assemble $10^9~M_{\odot}/h$ SMBHs by $z=6$. While these massive BHs are active as luminous quasars with near-Eddington luminosities of  $\sim10^{47}~\mathrm{ergs~s^{-1}}$ at $z=6$, their growth at $z\gtrsim 9$ is dominated by BH mergers. Overall, the BH growth is fastest within rare massive high-z halos that are also sufficiently compact and have low tidal fields. To produce $z\sim6$ quasars in these halos with the TNG accretion model, an early boost in BH mass driven by mergers is necessary.}
\label{bh_growth_env_fig}
\end{figure*}
\begin{figure*}
\includegraphics[width=14 cm]{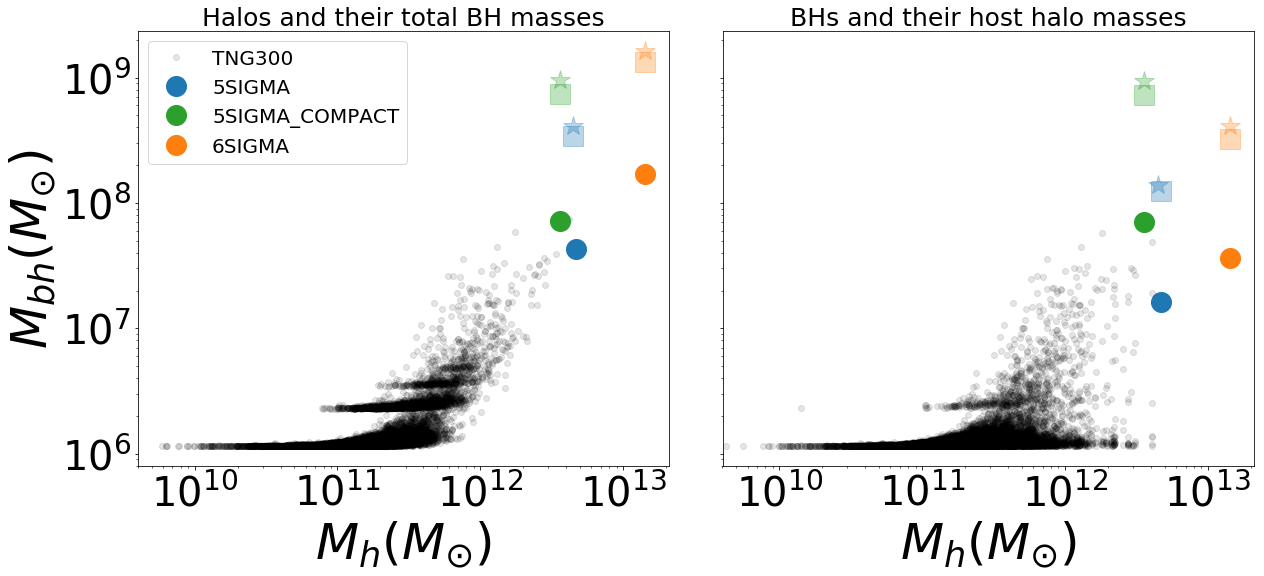}
\caption{BH mass vs. halo mass scaling relation at $z=6$. In the left panel, each data point corresponds to a halo; so we are plotting the ``halo mass vs. total mass of all BHs in the halo'', for all halos at $z=6$. In the right panel, each data point corresponds to a BH, so we are plotting the ``BH mass vs. host halo mass'', for all individual BHs at $z=6$. The blue, green and orange colors correspond to the BH within the MMH of \texttt{5SIGMA}, \texttt{5SIGMA_COMPACT} and \texttt{6SIGMA} regions respectively. All the runs use the TNG accretion model. The circles correspond to the TNG seed model, which we directly compare to the full population of TNG300.  The left panels show that for the TNG seed model, the 
total BH masses of 
MMHs produced by the constrained runs are 
consistent with the extrapolated trends from the TNG300 population.  This further validates our constrained runs. The squares $\&$ stars correspond to the gas-based seed model \texttt{SM5_FOF3000} with seed masses of $8\times10^5~\&~1\times10^5~M_{\odot}/h$, respectively; these models produce much higher BH masses 
compared to the TNG seed model. The MMHs generally tend to host a significantly large number of BHs during their assembly history, generally ranging between $\sim10-50$ depending on the constrained region as well as the seed model~(revisit Figure \ref{No_of_BHs_fig}). 
 The \texttt{6SIGMA} run produces the highest number of BHs and also the highest total BH mass in its target halo, commensurate with its halo mass. However, when we look at masses of \textit{individual} BHs in the right panel, the most massive BH is actually produced by \texttt{5SIGMA_COMPACT}. In fact, with the \texttt{SM5_FOF3000} seed model, only the \texttt{5SIGMA_COMPACT} can produce \textit{individual} $\sim10^9~M_{\odot}/h$ BHs similar to the observed $z\sim6$ quasars.}
\label{SM_BHM_relations_fig}
\end{figure*}
We are particularly interested in the growth histories of BHs occupying the MMH in each region. Figure \ref{No_of_BHs_fig} shows the number of BHs that are present within the MMH at different redshift snapshots, with different panels showing different seed models. We note that these MMHs tend to acquire several BHs during their assembly history. Despite many of these BHs inevitably merging with the central BH, the overall number of BHs hosted by the MMHs increases up to at least $z\sim9-10$. This is because the MMHs continue to acquire new BHs from surrounding merging halos. At $z=6$, the MMHs can generally host up to $\sim10-50$ BHs depending on the constrained region as well as the seed model.

Next, we look at how the number of BHs in the MMHs vary between different regions and seed models. For all seed models, we generally see that the \texttt{5SIGMA_COMPACT} runs tend to start forming seeds at earlier times, and therefore host a higher number of BHs at $z\sim10-20$ compared to \texttt{5SIGMA} and \texttt{6SIGMA} runs. This is because the MMHs at $z\sim10-20$ for the \texttt{5SIGMA_COMPACT} runs are more massive compared to the other two regions~(revisit Figure \ref{halo_evolution_fig}). However, between $z\sim10-6$, we find that there is no significant increase in the number of BHs for the MMH in \texttt{5SIGMA_COMPACT}, unlike \texttt{5SIGMA} and \texttt{6SIGMA}. This is likely because in the \texttt{5SIGMA_COMPACT} runs, most of the nearby massive halos have already merged with the MMH by $z\sim10$, leaving behind an isolated MMH during $z\sim10-6$ with very few nearby halos to acquire new BHs from. On the other hand, for the \texttt{5SIGMA} and \texttt{6SIGMA} runs, the MMHs are not as isolated at $z\sim10-6$, and they continue to acquire new BHs during this time. All of this ultimately leads to fewer BHs in the MMH of \texttt{5SIGMA_COMPACT} at $z=6$, compared to \texttt{5SIGMA} and \texttt{6SIGMA} regions. Lastly, note that the TNG seed model~(Figure \ref{No_of_BHs_fig}: left panel) is substantially more restrictive and produces much fewer seeds compared to our fiducial gas-based seed models with $\tilde{M}_{\mathrm{sf,mp}}=5, \tilde{M}_{\mathrm{h}}=3000$~(hereafter \texttt{SM5_FOF3000} shown in Figure \ref{No_of_BHs_fig}: middle and right panels). 

The primary science focus of this work is the growth of the most massive BH located in the MMH of our simulations. So unless otherwise stated, all future references to BH growth histories are for the most massive BH at $z=6$ in each simulation. Figure \ref{bh_growth_env_fig} shows the BH growth histories for \texttt{5SIGMA}, \texttt{5SIGMA_COMPACT} and \texttt{6SIGMA}. We first focus on the halo based TNG seed model~(leftmost panels), which starts to seed BHs around $z\sim10-12$. Note that for this seed model, very few BHs are formed overall, which results in very little growth via mergers~(see dashed lines in Figure \ref{bh_growth_env_fig}: top left panel). We find that amongst all three regions, \texttt{5SIGMA_COMPACT} assembles the highest mass BH at $z=6$, despite containing the least number of BHs within its MMH~(also recall that \texttt{6SIGMA} has a higher mass MMH at $z=6$). This is because the \texttt{5SIGMA_COMPACT} run produces a higher gas density at the peak location, thereby leading to the fastest growth via gas accretion at $z\lesssim9$. This result is overall consistent with N21, showing that compact peaks with low tidal fields are the most ideal environments for rapid BH growth. However, with this TNG seed model, the overall BH mass assembled at $z=6$ is only $\sim5\times10^{7}~M_{\odot}$, which is significantly smaller than the typical masses of the observed high-z quasars~($\sim10^9~M_{\odot}$).

Next, we look at the predictions from gas-based seed models, particularly \texttt{SM5_FOF3000}~($\tilde{M}_{\mathrm{sf,mp}}=5$ and $\tilde{M}_{\mathrm{h}}=3000$) with seed masses of $8\times10^5$ and $1\times10^5~M_{\odot}/h$~(Figure \ref{bh_growth_env_fig}: middle and right panels respectively). Again, these models produce substantially higher numbers of seeds that start forming at much earlier times~($z\sim17-25$) compared to the TNG seed model. As a result, there is now considerable growth via mergers. To that end, we note that regardless of how early the seeds form, accretion-driven BH growth does not become significant until $z\lesssim9$~(see 2nd rows of Figure \ref{bh_growth_env_fig}). This results from 
the fact that the central gas densities remain relatively low until $z\sim9$ but start to steeply increase between $z\sim9-6$~(revisit Figure \ref{1D_density_distributions_fig}). As a result, the BH growth at $z\gtrsim9$ is completely driven by BH mergers. In fact, for \texttt{SM5_FOF3000} seed models, the $z\gtrsim9$ merger-driven growth assembles a BH mass of $\sim3\times10^7~M_{\odot}$ by $z\sim9$, in contrast to the TNG seed model where the BHs are still close to the seed mass of $\sim10^6~M_{\odot}$ at $z\sim9$. Between $z\sim9-6$, the accretion-driven BH growth becomes increasingly significant for the \texttt{SM5_FOF3000} seed models, pushing 
the BH mass to values $\gtrsim 10^8~M_{\odot}$ at $z=6$ for all three constrained regions. Amongst the three regions, \texttt{5SIGMA_COMPACT} again produces the highest BH mass that now reaches close to $\sim10^9~M_{\odot}$ at $z=6$, consistent with the observed $z\sim6$ quasars. Additionally, note that the merger-driven growth can also be boosted, by simply reducing the halo mass threshold and forming more seeds. Therefore, $z\sim6$ quasars could also be formed within a ``halo mass only" seed model~(e.g. TNG seed model) with a sufficiently low halo mass threshold.  

The bolometric luminosities and Eddington ratios~(see 3rd and 4th rows of Figure \ref{bh_growth_env_fig}) of the BHs remain low~($L_{\mathrm{bol}}\sim10^{42}~\mathrm{ergs~s^{-1}}\sim10^{-3}~\mathrm{L_{\mathrm{bol}}^{\mathrm{edd}}}$) at $z\gtrsim9$ wherein the accretion-driven BH growth is insignificant. This is generally true for all seed models and constrained regions. As we go from $z\sim9-6$ during which the central gas densities steeply rise~(revisit Figure \ref{1D_density_distributions_fig}), the accretion-driven BH growth becomes increasingly efficient. This leads to a sharp increase in the BH luminosities. By $z\sim6$, the BHs start to grow close to the Eddington limit for all of the runs, generally corresponding to bolometric luminosities $\gtrsim10^{45}~\mathrm{ergs~s^{-1}}$. However, luminosities of observed $z\sim6$ quasars are even higher i.e. $\sim10^{47}~\mathrm{ergs~s^{-1}}$. These luminosities are produced only by the $\sim10^9~M_{\odot}$ BHs that are formed within $\texttt{5SIGMA_COMPACT}$ region using the \texttt{SM5_FOF3000} seed models. Overall, we find that to form BHs that resemble the observed $z\sim6$ quasars~(masses of $\sim10^9~M_{\odot}$ and luminosities of $\sim10^{47}~\mathrm{ergs~s^{-1}}$) in our simulations with \texttt{IllustrisTNG} physics, we need massive compact halos with seed models such as \texttt{SM5_FOF3000} that allow for substantial merger-driven BH growth at $z\gtrsim9$.

Figure \ref{SM_BHM_relations_fig} shows the predictions of our constrained runs 
the $z=6$ halo, as well as the BH populations in the TNG300 uniform simulation. Note that the most massive $z=6$ BHs produced by TNG300 are $\sim50$ times smaller than the observed $z\gtrsim6$ quasars. This is simply because TNG300 does not have the volume to produce such rare objects, despite being among the largest simulations to be run past $z=6$ and beyond. In fact, this is true for almost all major cosmological hydrodynamic simulations run to date ~(see \citealt{2021MNRAS.503.1940H,2022MNRAS.tmp..271H,2022MNRAS.509.3015H} for combined analyses of BH populations from several simulations). The only exception to this would be the \texttt{BlueTides} simulation~\citep{2016MNRAS.455.2778F} which produces a $6.4\times10^8~M_{\odot}$ BH in a volume of $[400~\mathrm{Mpc}/h]^3$ by $z\sim7.5$~\citep{2019MNRAS.483.1388T}. Overall, this further highlights the power of our constrained simulations which is able to produce such rare objects within smaller volume and higher resolution simulations in reasonable computing time, so as to allow an exploration of a wide range of model parameters.

We now specifically compare the BHs produced in the MMHs of the constrained runs to that of the TNG300 simulation. The left panel of Figure \ref{SM_BHM_relations_fig} shows the 
halo mass versus 
the total mass of all BHs within the halos. We find that for the TNG seed model, predictions for the constrained runs are consistent with extrapolation of the BH mass vs. halo mass relation from TNG300. 
These results, together with the stellar mass vs. halo mass relations in Figure \ref{SM_HM_relation_fig}, serve as a good validation for our constrained runs. As expected from the results in the previous paragraph, the gas-based seed models \texttt{SM5_FOF3000} produce halos with total BH masses that are significantly higher than the extrapolated trend of the TNG300 halos. As an additional note, the $M_{bh}-M_h$ relation for the lowest mass BHs within TNG300 form streaks of horizontal lines that can be clearly seen in Figure \ref{SM_BHM_relations_fig}. This is likely an artifact of the TNG seed model, where BHs seeded within $\gtrsim10^{10}~M_{\odot}$ halos do not show significant accretion-driven growth until halos reach masses $\gtrsim10^{11}~M_{\odot}$.   

Closer examination of Figure \ref{SM_BHM_relations_fig} (left panel) reveals another interesting result. For the TNG seed model as well as the gas-based seed model, the MMH in \texttt{6SIGMA} achieves the highest {\em total} BH mass. 
This contrasts with Figure \ref{bh_growth_env_fig} and with the right panel of Figure \ref{SM_BHM_relations_fig}, which clearly show that \texttt{5SIGMA_COMPACT} produces the highest {\em individual} BH mass in the MMH. 
As it turns out, while the MMHs in \texttt{5SIGMA} and \texttt{6SIGMA} end up with a higher number of BHs than \texttt{5SIGMA_COMPACT}~(revisit Figure \ref{No_of_BHs_fig}), the individual BHs in \texttt{5SIGMA} and \texttt{6SIGMA} are significantly smaller than the most massive BH in \texttt{5SIGMA_COMPACT}. The foregoing statement is generally true for TNG seed model as well as the gas-based seed models. Particularly for the gas-based seed model \texttt{SM5_FOF3000}, this 
means that while a typical rare massive halo~($\sim10^{13}~M_{\odot}$) at $z\sim6$ can acquire total BH mass exceeding $\sim10^9~M_{\odot}$, it may not produce ``individual BHs" of such masses. Therefore, to host the observable $z\gtrsim6$ quasars that correspond to individual $\sim10^9~M_{\odot}$ BHs growing close to the Eddington limit, we need halos that are not just massive enough, but are also highly compact and have low tidal fields. In these compact MMHs, the BHs are more likely to have close encounters with each other. Therefore they can merge more readily to form a single massive BH at the halo centers, compared to typical halos of the same mass. To that end, recall that the small scale dynamics are poorly resolved in our simulations, particularly for lower mass BHs.  Several recent works with more realistic treatment of BH small scale dynamics~\citep[for e.g.][]{2017MNRAS.470.1121T,2018ApJ...857L..22T,2022MNRAS.tmp..460N,2022MNRAS.510..531C} have found that many of the seeds~(particularly lower mass seeds) do not sink efficiently to the local potential minima, thereby leading to a population of wandering BHs~\citep{2018ApJ...857L..22T,2021MNRAS.503.6098R,2021ApJ...916L..18R,2021MNRAS.508.1973M,2022MNRAS.tmp..221W}. Therefore, prompt mergers resulting from our current BH repositioning scheme could overestimate the rate at which the central BH grows. In future work, we shall investigate this in the context of the assembly of the $z\sim6$ quasars. 


\begin{figure*}
\includegraphics[width=8 cm]{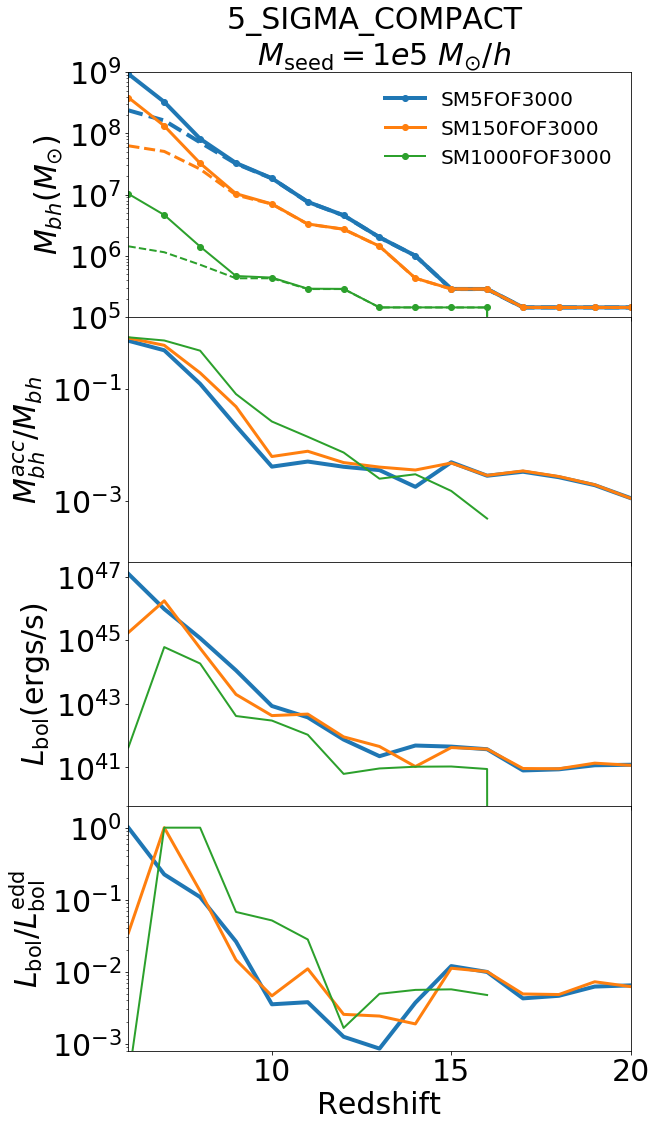}
\includegraphics[width=8 cm]{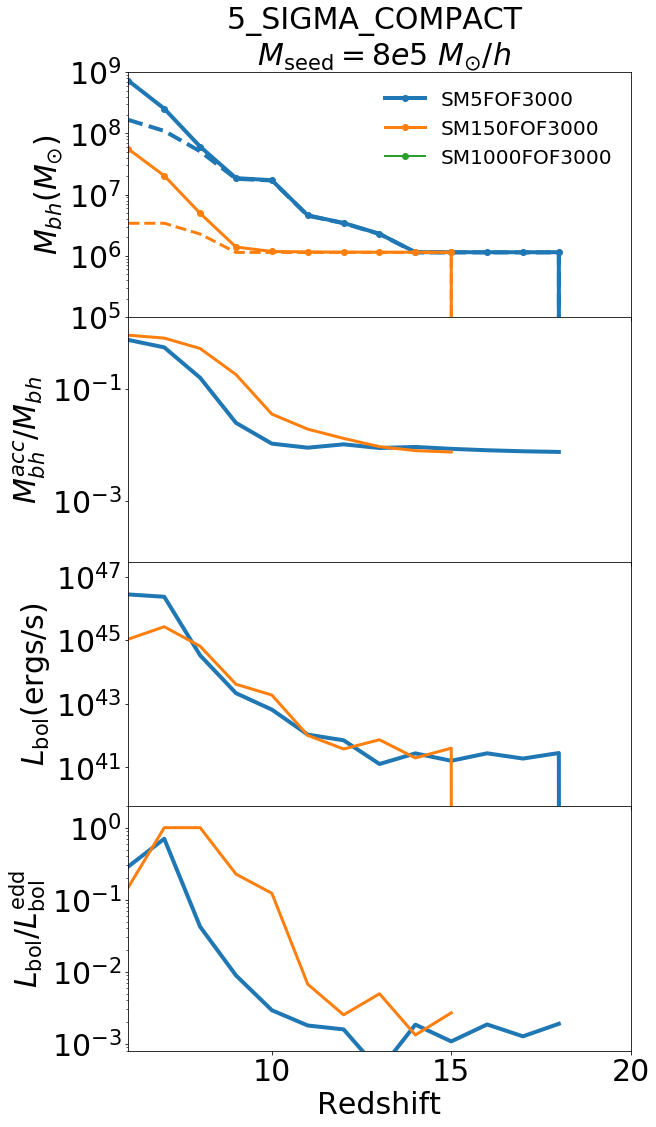}
\caption{Evolution of the most massive black hole in \texttt{5SIGMA_COMPACT} for $\tilde{M}_{\mathrm{sf,mp}}=5,150~\&~1000$. Left and right panels correspond to seed masses of $10^5~M_{\odot}/h$ and $8\times10^5~M_{\odot}/h$ respectively. All the runs here use the TNG accretion model. The different rows show the same set of quantities as Figure \ref{bh_growth_env_fig}. As the seeding criteria becomes more stringent, there are fewer mergers to grow the BH at $z\gtrsim9$. As a result, the final $z=6$ BH mass decreases and falls significantly short of producing the observed $z\sim6$ quasars. }
\label{gas_based_models_fig}
\end{figure*}

Next, we focus on the \texttt{5SIGMA_COMPACT} region and further explore different variations of gas-based seeding models to study their impact on BH growth. Note that \texttt{SM5_FOF3000}~($\tilde{M}_{\mathrm{sf,mp}}=5~\&~\tilde{M}_{\mathrm{h}}=3000$), which successfully produces a $z\sim6$ quasar, is the least restrictive amongst the family of BH seeding models developed in \cite{2021MNRAS.507.2012B,2021MNRAS.tmp.3110B}. Figure \ref{gas_based_models_fig} shows the impact of further increasing $\tilde{M}_{\mathrm{sf,mp}}$ to values of 150 and 1000, on the BH mass, luminosity and Eddington ratio evolution. As we increase $\tilde{M}_{\mathrm{sf,mp}}$, fewer seeds form and the merger-driven BH growth is commensurately suppressed. This leads to a significant slow-down of BH growth, and thereby decreases the BH mass assembled by $z=6$. For $10^5~M_{\odot}/h$ seeds, increasing $\tilde{M}_{\mathrm{sf,mp}}$ from $5$ to $1000$ decreases the final $z=6$ BH mass by a factor of $\sim100$. The $z=6$ luminosities also drop from $\sim10^{47}~\mathrm{erg~s^{-1}}$ to $\sim10^{43}~\mathrm{erg~s^{-1}}$. For more massive $8\times10^5~M_{\odot}/h$ seeds, the impact is significantly stronger~(no $8\times10^5~M_{\odot}/h$ seeds form for $\tilde{M}_{\mathrm{sf,mp}}=1000$). In general, any gas-based seeding prescription that is more restrictive than \texttt{SM5_FOF3000} fails to produce BHs consistent with the observed $z\sim6$ quasars.

Overall, we find that within the TNG galaxy formation model, the \texttt{SM5_FOF3000} gas-based seed model is able to successfully reproduce $z\sim6$ BHs that are comparable to the observed high-z quasars, but only in massive ($\sim10^{12}~M_{\odot}$) halos that are highly compact and have low tidal fields. Additionally, both 1) merger-dominated growth at $z\gtrsim9$, and 2) accretion-dominated growth at  $z\sim6-9$ are crucial for producing these high-z quasars. 

\subsection{Implications for strongly restrictive seed models in producing $z\gtrsim6$ quasars}
\label{Impact of BH accretion sec}

\begin{figure*}
\includegraphics[width=8 cm]{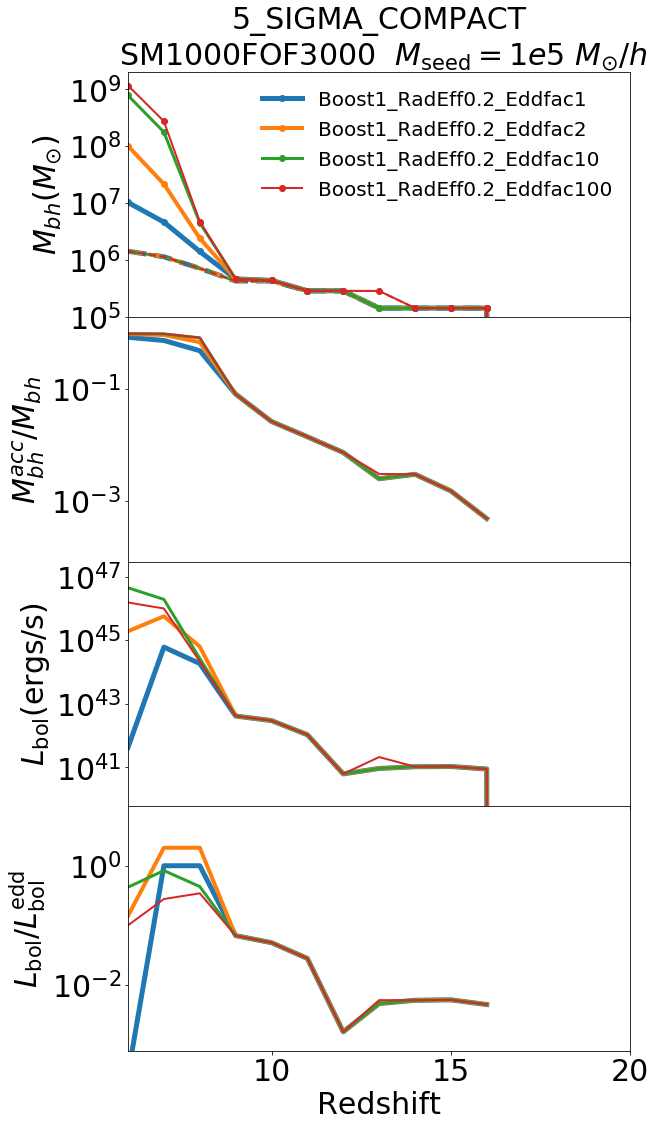}
\includegraphics[width=8 cm]{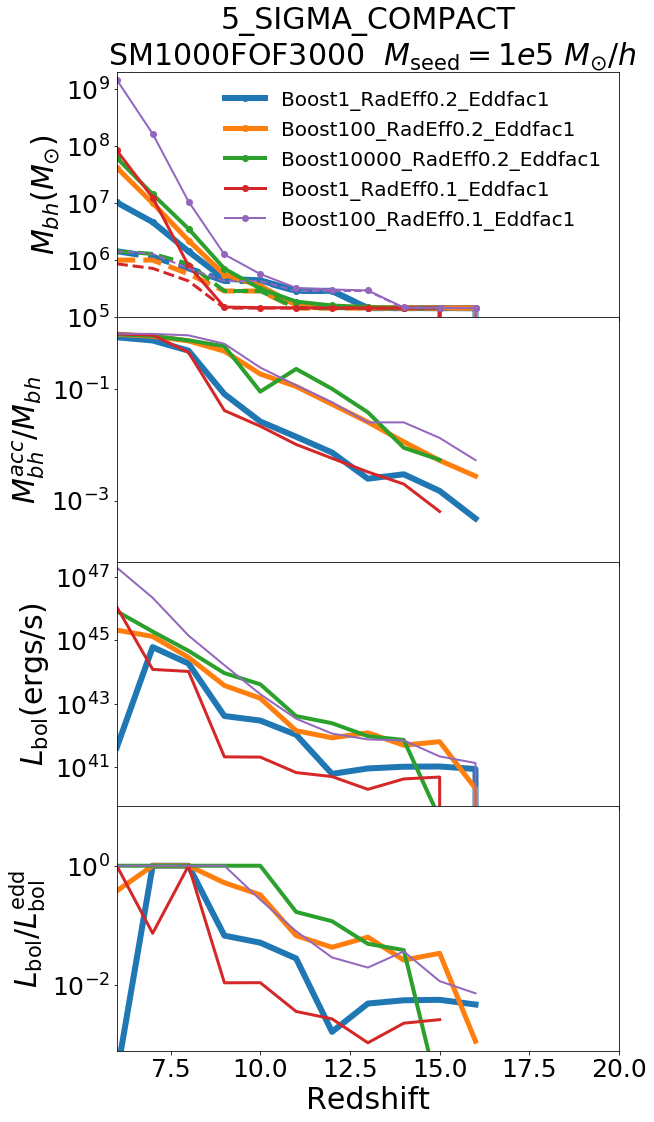}

\caption{Evolution of the most massive black hole in the \texttt{5SIGMA_COMPACT} volume for different accretion models labelled hereafter as \texttt{Boost*_RadEff*_EddFac*} where the `*'s correspond to values of $\alpha$, $\epsilon_r$ and $f_{\mathrm{edd}}$ respectively. All the runs use the most stringent gas-based seed model \texttt{SM1000_FOF3000}, wherein the growth via mergers is small. We then explore different accretion models. In the left panel, we keep $\alpha=1$ and $\epsilon_r=0.2$ fixed, and show the BH growth histories for different values of $f_{\mathrm{edd}}$ between 1-100. In the right panel, we consider different combinations of $\alpha=1-10000$ and $\epsilon_r=0.2~\&~0.1$. In the absence or lack of mergers, accretion alone can assemble a $10^9~M_{\odot}$ BH at $z\sim6$ only if we enhance the maximum allowed accretion rate compared to the TNG accretion model. This can be achieved by either allowing for super-Eddington accretion rate, or by reducing the radiative efficiency. In contrast, a higher Bondi boost alone does not sufficiently enhance BH growth to assemble a $z\sim6$ quasar.}
\label{varying_accretion_fig}
\end{figure*}

\begin{figure}
\includegraphics[width=8 cm]{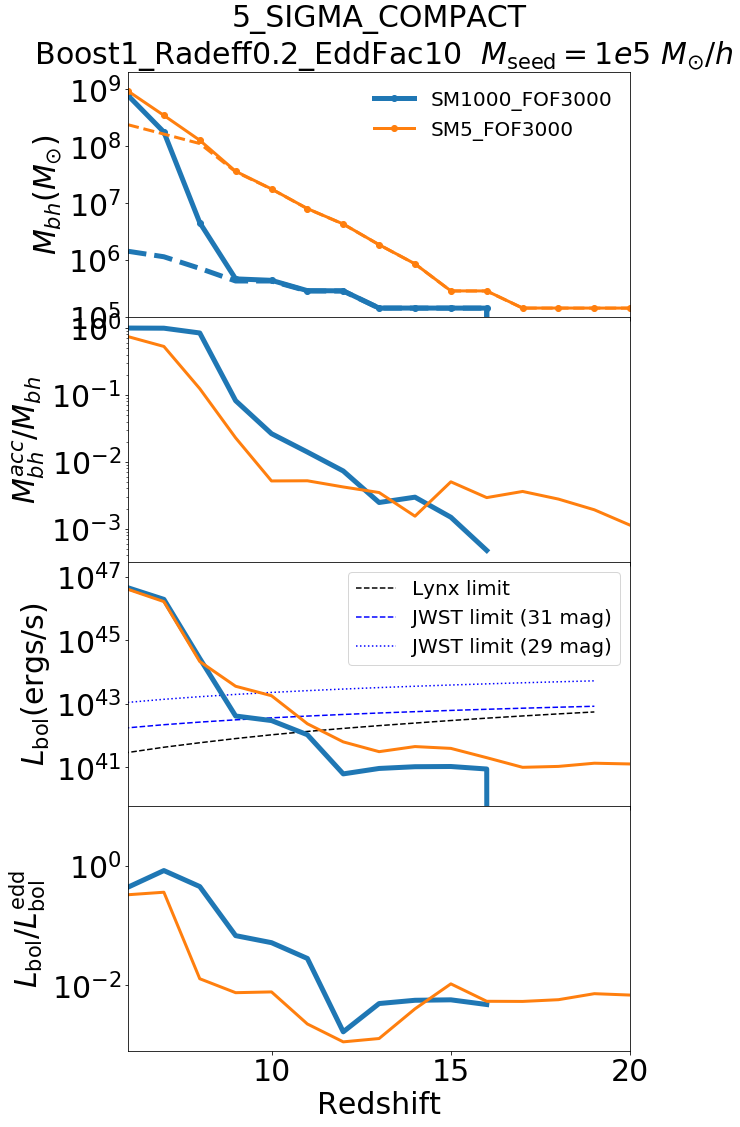}
\caption{Here we consider the model \texttt{Boost1_Radeff0.2_Eddfac10} that can already grow a $z\sim6$ quasar via accretion alone, and then investigate the impact of enhancing the number of seeds and mergers on the final BH mass at $z=6$. Blue lines correspond to $\tilde{M}_{\mathrm{sf,mp}}=1000$~(\texttt{SM1000_FOF3000}), which does not produce many seeds and mergers. Orange lines correspond to $\tilde{M}_{\mathrm{sf,mp}}=5$~(\texttt{SM5_FOF3000}) which does produce a substantial number of seeds and mergers. We find that in models such as  \texttt{Boost1_Radeff0.2_Eddfac10} wherein BHs can already grow to $\sim10^{9}~M_{\odot}$ via accretion alone, introducing more seeds and mergers~(by reducing $\tilde{M}_{\mathrm{sf,mp}}$) does not lead to any further increase in the final BH mass at $z=6$. In the 3rd panel, the black dashed line is the detection limit of $10^{-19}~\mathrm{ergs~cm^{-2}~s^{-1}}$ of the Lynx $2-10~\mathrm{keV}$ band, derived using bolometric corrections adopted from \protect\cite{2007MNRAS.381.1235V}. The blue dashed and dotted lines are JWST detection limits of 31 and 29th apparent magnitude for exposure times of $10^5$ and $10^4\mathrm{s}$ respectively~(same as those assumed in \protect\citealt{2020MNRAS.492.5167V}), with bolometric correction adopted from \protect\cite{1994ApJS...95....1E}. Therefore, JWST and Lynx observations of the quasar progenitors at $z\sim9-10$ could potentially contain signatures of their seeding environments.} 
\label{varying_accretion_edd}
\end{figure}

We have thus far seen that $z\sim6$ quasars cannot be assembled by our constrained simulations without an early boost in BH mass via mergers. This can only occur for relatively less restrictive seed models~($\tilde{M}_{\mathrm{sf,mp}}=5$), wherein enough seeds are formed to substantially contribute to merger-driven BH growth. Here we look for circumstances under which more restrictive seed models can produce $z\sim6$ quasars even in the absence of sufficient mergers. In particular, we explore models with enhanced accretion in these highly biased regions, compared to the TNG accretion model.

In Figure \ref{varying_accretion_fig}, we take one of our most restrictive seeding models i.e. $\tilde{M}_{\mathrm{sf,mp}}=1000$, and explore different accretion models to identify the ones that can produce $z\sim6$ quasars. As already noted in the previous section, only a handful of seeding and merger events occur for this model. We then investigate the BH growth under different variations for the accretion model. In the left panel, we keep $\alpha=1~\&~\epsilon_r=0.2$ fixed and vary $f_{\mathrm{edd}}$ from $1-100$~(recall that  sets the maximum accretion rate in the units of the Eddington rate). Not unexpectedly, we find that as $f_{\mathrm{edd}}$ is increased, the final BH mass is enhanced via accretion-driven BH growth. $f_{\mathrm{edd}}\gtrsim10$ is required for growing BHs to $\sim10^9~M_{\odot}$ with luminosities $\sim10^{47}~\mathrm{erg~s^{-1}}$ by $z=6$~(Figure \ref{varying_accretion_fig}: green line).  Notably, further increasing $f_{\mathrm{edd}}$ to 100 does not lead to any substantial increase in the final BH mass beyond $\sim10^9~M_{\odot}$~(Figure \ref{varying_accretion_fig}, right panel: red line). This is because the Bondi accretion rate exceeds $100~\times$ the Eddington limit for only a small fraction of the time.

We now keep $f_{\mathrm{edd}}$ fixed at 1 and examine the impact of the boost factor $\alpha$. When $\alpha$ is increased from $1$ to $100$, it leads to a factor $\sim5$ increase in the final BH mass at $z=6$~(blue vs orange lines in Figure \ref{varying_accretion_fig}: right panel). This is substantially smaller than the impact of increasing $f_{\mathrm{edd}}$ to 10. Additionally, further increasing the boost factor to 10000~(green lines in Figure \ref{varying_accretion_fig}: right panel) makes no significant difference in the $z=6$ BH mass. This implies that the maximum accretion rate set by the Eddington factor is much more consequential to the $z=6$ BH mass, compared to the Bondi boost factor. This is because the majority of the BH mass assembly occurs at $z\lesssim9$ when the accretion rates are already at their maximum allowed value. To that end, note that the maximum accretion rate can also be increased by decreasing the radiative efficiency $\epsilon_r$. Several cosmological simulations~(including N21) use a lower efficiency of $\epsilon_r=0.1$. If we fix $\alpha=1$ and decrease the radiative efficiency from $0.2$ to $0.1$, the $z=6$ BH mass increases by factor of $10$~(see blue vs red lines in Figure \ref{varying_accretion_fig}: right panel). Not surprisingly, this is similar to what we found when $f_{\mathrm{edd}}$ was increased to 2~(revisit Figure \ref{varying_accretion_fig}: left panel). Notably, at this lower radiative efficiency of $\epsilon_r=0.1$, applying a boost of $\alpha=100$ forms a $\sim10^9~M_{\odot}$ BH with luminosity of $\sim10^{47}~\mathrm{erg~s^{-1}}$ at $z=6$~(purple line in Figure \ref{varying_accretion_fig}: right panel). This is consistent with the results of N21, as we shall see in more detail in Section \ref{Comparison with other theoretical works}.  
To summarize our results thus far, we have shown that to form the observed $z\sim6$ quasars within rare dense compact halos in our simulations, one of the following two requirements must be fulfilled: \begin{enumerate} \item When the default TNG model is used (i.e., $\alpha=1,\epsilon_r=0.2~\&~f_{\mathrm{edd}}=1$), we need a sufficiently early~($z\gtrsim9$) boost in BH mass driven by BH mergers to grow to $\sim10^9~M_{\odot}$ by $z=6$. Our gas-based seeding prescription with $\tilde{M}_{\mathrm{sf,mp}}=5$ and $\tilde{M}_h=3000$ satisfies this requirement. \item For more restrictive seeding models~($\tilde{M}_{\mathrm{sf,mp}}\gtrsim150$) where mergers are absent or insufficient, $z\sim6$ quasars cannot be produced unless we enhance the maximum allowed accretion rate~(by factors $\gtrsim10$) within these extreme overdense regions by either increasing the Eddington factor or decreasing the radiative efficiency. Notably, this result is consistent with the recent work of \citep{2022arXiv220412513H} which uses semi-analytic approach to produce $z\sim6$ quasars using super-Eddington accretion, from both light~($10~M_{\odot}$) and heavy seeds~($10^5~M_{\odot}$).  
\end{enumerate}

\subsection{Impact of seed model on BH growth for `growth optimized' accretion parameters}
Given that some of these BH accretion models can produce massive BHs by $z\sim6$ without an early boost from BH mergers, we now explore what happens when these `growth optimized' accretion parameters are combined with merger-driven growth from less restrictive BH seed models. In the left panel of Figure \ref{varying_accretion_edd}, we take the accretion model $\alpha=1, \epsilon_r=0.2~\&~f_{\mathrm{edd}}=10$, and compare the BH growth histories for two seed models with $\tilde{M}_{\mathrm{sf,mp}}=5~\&~1000$~(\texttt{SM5_FOF3000} \& \texttt{SM1000_FOF3000} respectively). We have already seen that \texttt{SM1000_FOF3000} produces a $\sim10^9~M_{\odot}$ BH with  $\sim10^{47}~\mathrm{erg~s^{-1}}$ luminosity at $z=6$ even in the absence of significant number of mergers. We now examine whether the substantial merger-driven growth of \texttt{SM5_FOF3000} leads to any further increase in the $z=6$ BH mass much beyond $\sim10^9~M_{\odot}$, when combined with super-Eddington accretion of $f_{\mathrm{edd}}=10$. The \texttt{SM5_FOF3000} model grows BHs via mergers to $\sim10^7~M_{\odot}$ by $z\sim9$, which is a factor of $\sim 500$ higher than the \texttt{SM1000_FOF3000} model. Despite this large difference in masses, the luminosities are similar in both models, such that the higher-mass BH in \texttt{SM5_FOF3000} has a lower Eddington ratio. By
 $z=6$, the BH in \texttt{SM5_FOF3000}
reaches a mass of $\sim10^9~M_{\odot}$, similar to \texttt{SM1000_FOF3000}. To summarize, if a given model is already producing a $\sim10^9~\mathrm{M_{\odot}}$ BH at $z\sim6$ via accretion alone, boosting the merger-driven BH growth by forming more seeds does not further increase the final $z=6$ BH mass by a significant amount. 

The results from Figure \ref{varying_accretion_edd} also imply that if the accretion model is such that $z\sim6$ quasars can be assembled via accretion alone, multiple sets of seed models can produce the observed $z\sim6$ quasars. In such a case, the $z\sim6$ quasar observations alone may not be able to constrain BH seed models. However, the progenitors of these quasars at $z\gtrsim9$ can have significantly different assembly histories depending on the seed model, particularly in terms of the contribution from BH mergers. In fact, \texttt{SM5_FOF3000} naturally predicts a $\sim100$ times higher number of mergers compared to \texttt{SM1000_FOF3000}. Moreover, these merging progenitors will likely include the most massive black holes at their respective redshift. Therefore, detection of the loudest LISA events at $z\gtrsim9$ are likely to provide strong constraints for seed models. In terms of electromagnetic observations, the AGN progenitors are above of the detection limits of Lynx and JWST~(with limiting apparent magnitude of 31) up to $z\sim10$; this is true for both \texttt{SM5_FOF3000} and \texttt{SM1000_FOF3000} models~(revisit 3rd row of Figure \ref{varying_accretion_edd}). However, the difference in luminosities produced by both seed models is within a factor of $\sim10$, corresponding to a magnitude difference of only $\sim2.5$. Therefore, it is likely going to be difficult to find imprints of seed models within Lynx and JWST observations of the brightest AGN at higher redshifts~($z\gtrsim9$). 

\subsection{BH growth at higher resolutions}
\label{BH growth at higher resolutions_sec}
\begin{figure}
\includegraphics[width=8.5 cm]{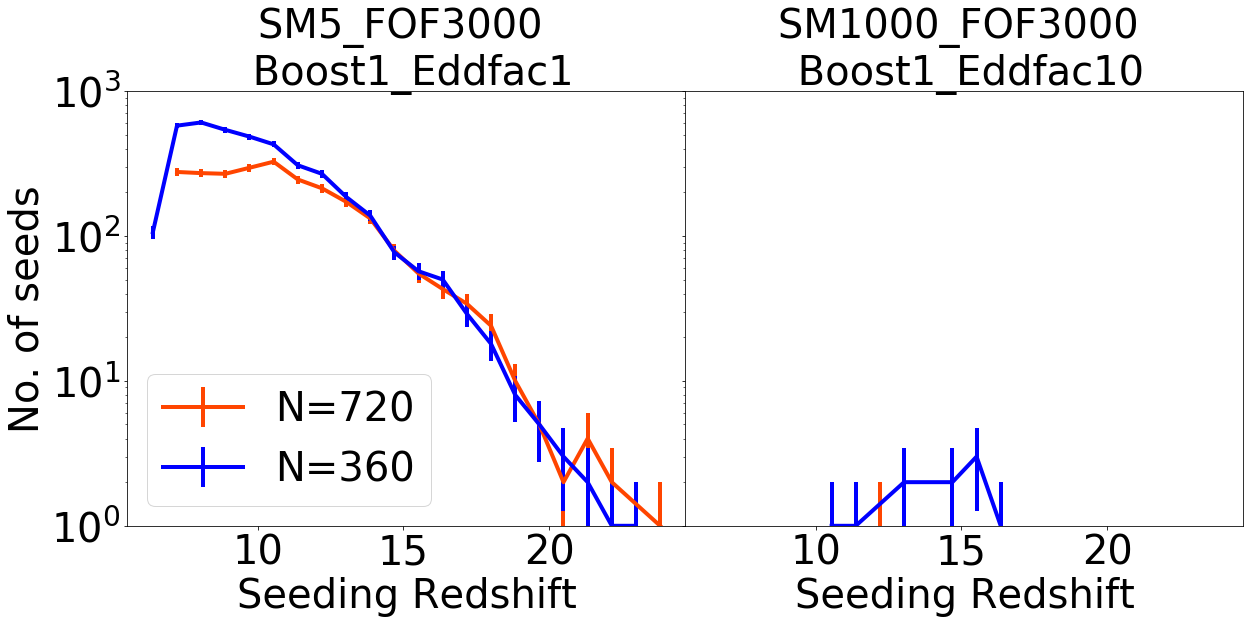}
\includegraphics[width=8.5 cm]{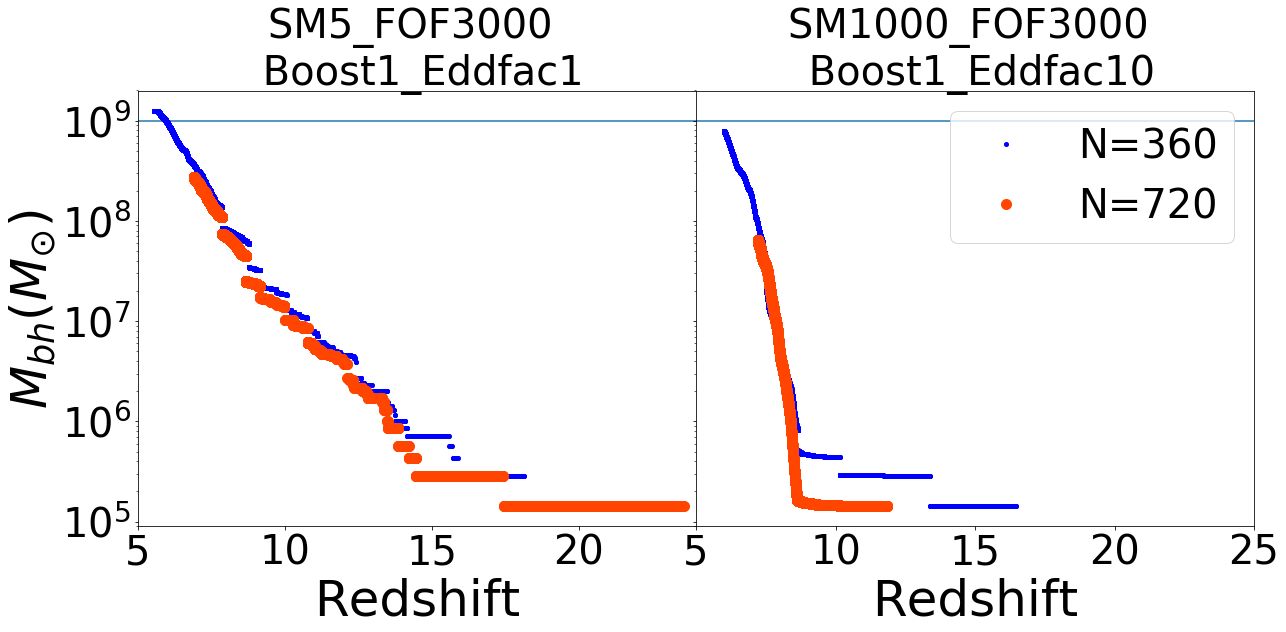}
\caption{Top and bottom panels show the total number of seeds formed and 
growth of the most massive BH respectively for $M_{\mathrm{seed}}=10^5~M_{\odot}/h$ at two different resolutions, namely $N=360$ and $N=720$. We show the resolution convergence for two distinct seed models which produce a $\sim10^9~M_{\odot}$ BH by $z=6$. Left panels correspond to the least restrictive seeding model of $\tilde{M}_{\mathrm{sf,mp}}=5, \tilde{M}_{\mathrm{h}}=3000$, wherein mergers substantially dominate the growth at $z\gtrsim9$ and accretion model is Eddington limited with $\alpha=1,\epsilon_r=0.2,f_{\mathrm{edd}}=1$. The right panels correspond to the much more restrictive seed model of $\tilde{M}_{\mathrm{sf,mp}}=1000, \tilde{M}_{\mathrm{h}}=3000$ with insignificant merger-driven growth; the accretion model corresponds to $\alpha=1,\epsilon_r=0.2,f_{\mathrm{edd}}=10$. In both the seed models, the final BH masses at $z\sim7$ are similar for $N=360$ and $N=720$.}
\label{resolution_convergence_fig}
\end{figure}
We have thus far presented results at our fiducial resolution of $N=360$, corresponding to gas mass resolution of $\sim10^5~M_{\odot}/h$. While we were able explore a wide range of models at this resolution at reasonable computational cost, we demonstrated in \cite{2021MNRAS.507.2012B} that our gas-based seed models start to become reasonably well converged only at resolutions $\lesssim10^4~M_{\odot}/h$. Therefore, it is imperative to perform a resolution convergence test by running some of these simulations at gas mass resolutions of $\sim10^4~M_{\odot}/h$~($N=720$). Particularly, we consider the seed models \texttt{SM5_FOF3000} with $\alpha=1,\epsilon_r=0.2,f_{\mathrm{edd}}=1$, and \texttt{SM1000_FOF3000} with $\alpha=1,\epsilon_r=0.2,f_{\mathrm{edd}}=10$, both of which successfully produced a $\sim10^9~M_{\odot}$ quasar by $z\sim6$. Due to computational reasons, we could only run the higher resolution simulations~($N=720$) to $z=7$. 

The results are shown in Figure \ref{resolution_convergence_fig}, where they are compared to the lower resolution runs~($N=360$). Let us start with \texttt{SM5_FOF3000}~(left panels), which produces enough seeds to allow for substantial amounts of merger-driven BH growth. The number of seeds formed~(Figure \ref{resolution_convergence_fig}: top left panel) is similar between $N=360~\&~720$ for $z\gtrsim12$. As shown in \cite{2021MNRAS.507.2012B}, at these redshifts, seeding is largely driven by the proliferation of new star forming regions~(star formation is reasonably well converged between gas mass resolutions of $\lesssim10^5~M_{\odot}/h$). At $z\lesssim12$, the higher resolution simulation produces a somewhat lower number of seeds~(by factors up to $\sim5$). The slower resolution convergence at $z\sim7-12$ is also fully consistent with the zoom simulations of \cite{2021MNRAS.507.2012B}. It is due to the markedly stronger metal dispersion for higher resolutions, which causes a stronger suppression of seeding at $z\sim7-12$ relative to lower resolution simulations. Nevertheless, the final $z=7$ BH mass of $\sim10^8~M_{\odot}/h$ assembled by the higher resolution simulation~(Figure \ref{resolution_convergence_fig}: bottom left panel), is only slightly smaller~(by a factor of $\sim1.5$) compared to the lower resolution simulation. This strongly indicates at even at higher resolutions, the \texttt{SM5_FOF3000} seed model would be able to assemble a $\sim10^9~M_{\odot}$ by $z=6$. 

Now let us focus on the \texttt{SM1000_FOF3000} model~(with $f_{\mathrm{edd}}=10$), where the merger-driven BH growth is minimal and super-Eddington growth is used to produce a $\sim10^9~M_{\odot}$ by $z=6$. Here, the higher resolution run~($N=720$) produces only 1 seed, whereas the lower resolution produced $\sim10$ seeds~(Figure \ref{resolution_convergence_fig}: top right panel). This is also consistent with our findings in \cite{2021MNRAS.507.2012B} where we showed that resolution convergence becomes poorer as seed models become more restrictive with higher $\tilde{M}_{\mathrm{sf,mp}}$. Despite this, the accretion-driven BH growth at $z\lesssim9$ assembles a BH mass close to $\sim10^8~M_{\odot}$ by $z\sim7$ for both $N=720~\&~360$ resolutions~(Figure \ref{resolution_convergence_fig}: bottom right panel). Overall, we find that the BH models which successfully produce a $z\gtrsim6$ quasar at our fiducial resolution~($N=360$), will likely continue to do so at even higher resolutions.     
\subsection{DCBHs as possible seeds of $z\gtrsim7$ quasars}

\begin{figure}
\includegraphics[width=8 cm]{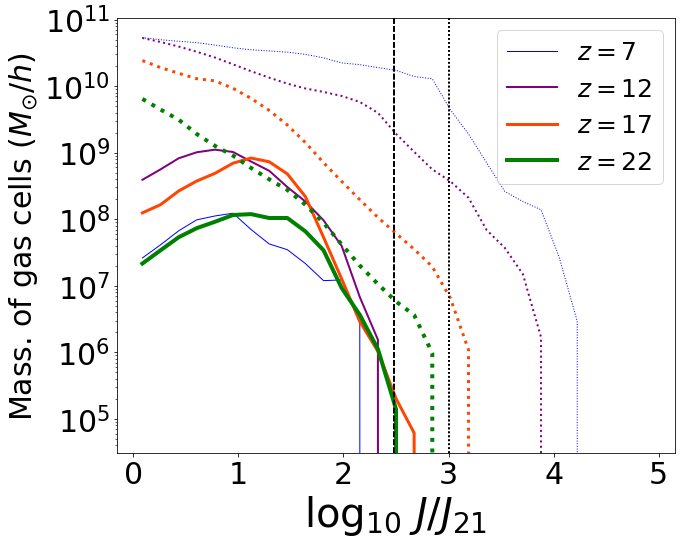}
\caption{Total mass of gas cells illuminated by LW photons originating from Pop II~($0.001<Z<0.1~Z_{\odot}$) and Pop III~($Z<0.001~Z_{\odot}$) stars within bins of various flux values shown in the x-axis. These are runs at $N=720$~(higher than the fiducial resolution). Dotted lines correspond to all gas cells and solid lines correspond to dense, metal poor gas cells. Black vertical lines correspond to flux thresholds of $J_{\mathrm{crit}}=300~J_{21}$~(dashed) and $J_{\mathrm{crit}}=1000~J_{21}$~(dashed). When we look at all gas cells, flux values can reach up to $\sim10^4~J_{21}$. However, within dense, metal poor regions, flux values reach only up to $\sim300~J_{21}$. Therefore, for direct collapse black hole~(DCBHs) seeds to form in our constrained regions, the critical LW flux values need to be less than $\sim300~J_{21}$. One-zone chemistry models and high resolution hydrodynamic simulations predict significantly higher critical flux~($\gtrsim1000~J_{21}$).}
\label{LW_distribution_fig}
\end{figure}

Here we use the higher resolution simulations~($N=720$) to investigate the possibility of DCBHs as candidates for the seeds of the $z\gtrsim6$ quasars. The conditions for their formation are very restrictive due to the requirement of high LW flux incident upon dense, pristine gas. In particular, small scale hydrodynamic simulations~\citep{2010MNRAS.402.1249S} and one-zone chemistry models~\citep{2014MNRAS.445..544S,2017MNRAS.469.3329W} infer critical LW fluxes to be $\gtrsim10^3~J_{21}$. In the zoom simulations of \citep{2021MNRAS.tmp.3110B} that contained a $3.3\sigma$ overdense peak~(for a target halo mass of $3.5\times10^{11}~M_{\odot}/h$ at $z=5$), the highest LW flux incident within dense, metal poor gas was $\sim100~J_{21}$. The work showed that the bulk of the $z\gtrsim6$ SMBH population may be difficult to explain via the DCBH seeding channel. 

We investigate whether DCBHs can form in the \texttt{5SIGMA_COMPACT} region, which is much more extreme compared to the region probed in \cite{2021MNRAS.tmp.3110B}. Figure \ref{LW_distribution_fig} shows the distribution of LW intensities at various redshifts, across all gas cells~(dotted lines) as well as dense, metal poor gas cells~(solid lines). In general, the LW intensities can be as high as $\sim10^4~J_{21}$. However, when we exclusively look at dense, metal poor gas, the LW intensities span only up to $\sim300~J_{21}$. When we apply a critical flux threshold of $\sim300~J_{21}$ for seeding~(\texttt{SM5_FOF3000_LW300}), only a handful of seeds are formed~(Figure \ref{seedmass_dependence}: upper right panel). For higher critical fluxes e.g. $\sim10^3~J_{21}$, there is no DCBH seed formation. Therefore, our model can support DCBHs as potential candidates for the seeds of $z\gtrsim6$ quasars only if the critical LW flux is $\lesssim300~J_{21}$. Secondly, due to the absence of any significant merger-driven BH growth, to produce $z\sim6$ quasars from these very few DCBH seeds, we need to enhance the maximum allowed accretion rate compared to the TNG accretion model by allowing for super-Eddington accretion or reducing the radiative efficiency~(as follows from the results of Section \ref{Impact of BH accretion sec}). As an example, the lower right panel of Figure \ref{seedmass_dependence} shows the growth of a DCBH seed~($J_{\mathrm{crit}}=300~J_{21}$) to $\sim10^8~M_{\odot}/h$ by $z=7$ via super-Eddington accretion with $f_{\mathrm{edd}}=10$. Given the trends from the lower resolution $N=360$ simulations, we expect the BH to continue growing to $\sim10^9~M_{\odot}$ by $z=6$. 

\subsection{Impact of seed mass on the formation of $z\gtrsim6$ quasars}
\begin{figure}
\includegraphics[width=8.5 cm]{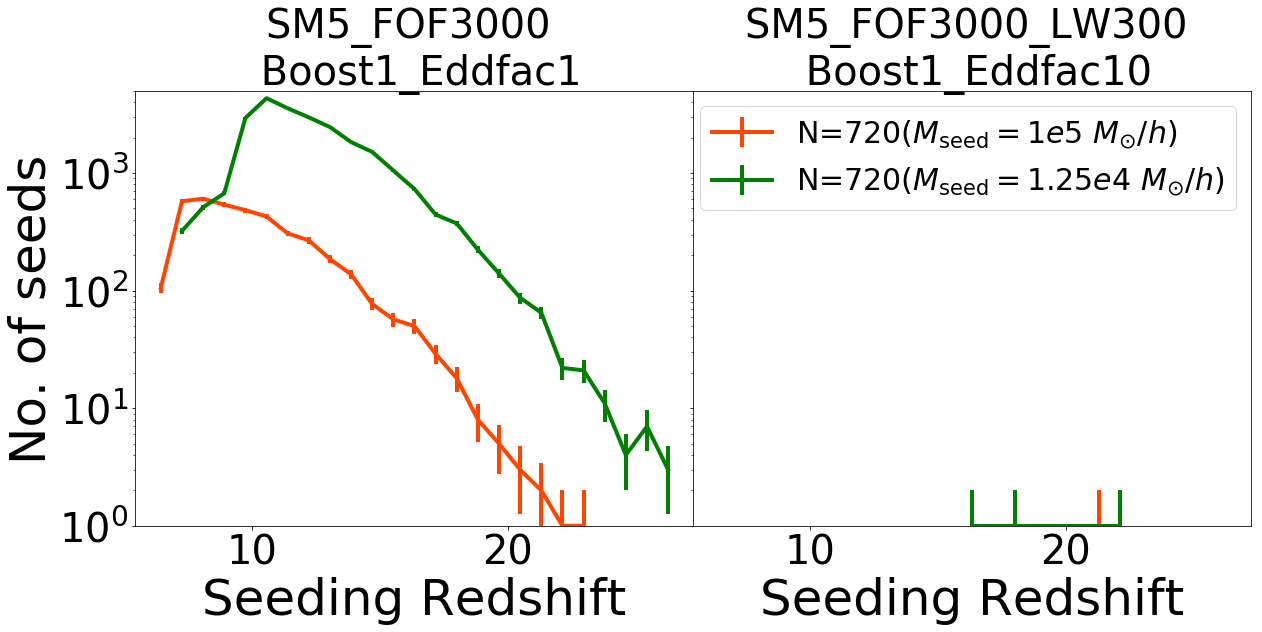}
\includegraphics[width=8.5 cm]{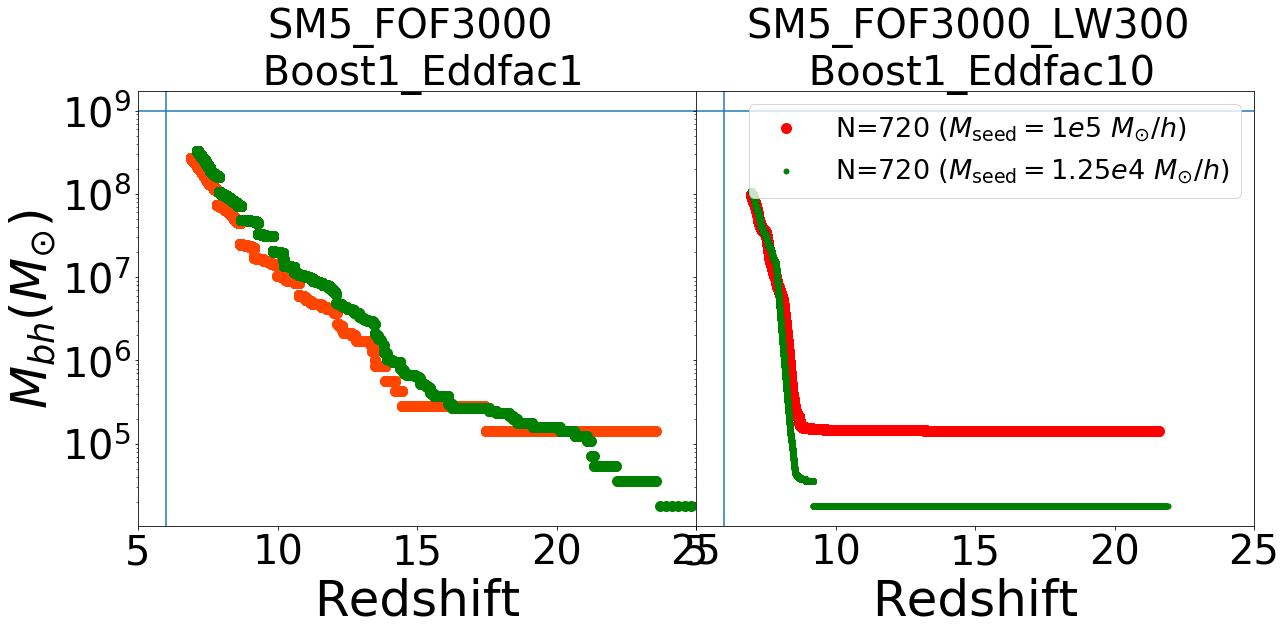}
\caption{Top and bottom panels, respectively, show the number of seeds and BH mass growth at $N=720$ for two different seed masses of $M_{\mathrm{seed}}=10^5~\&~1.25\times10^4~M_{\odot}/h$ . Left panels correspond to the least restrictive seeding model of $\tilde{M}_{\mathrm{sf,mp}}=5, \tilde{M}_{\mathrm{h}}=3000$, wherein mergers substantially dominate the growth at $z\gtrsim9$ and accretion rate is Eddington limited. Here, $1.25\times10^4~M_{\odot}/h$ seeds form and merge $\sim8$ times more frequently than $1\times10^5~M_{\odot}/h$ seeds. Both seed masses grow to similar mass BHs by $z\sim7$. The right panels correspond to the much more restrictive seed model of $\tilde{M}_{\mathrm{sf,mp}}=5, \tilde{M}_{\mathrm{h}}=3000$ and critical LW flux of $300~J_{21}$; the accretion model is given by  $\alpha=1,\epsilon_r=0.2,f_{\mathrm{edd}}=10$. Here, the merger-driven growth is insignificant. Regardless, both seed masses produce similar mass BHs by $z\sim7$.}
\label{seedmass_dependence}
\end{figure}

At our fiducial resolution~($N=360$), we were able to probe seed masses of $10^5~\&~8\times10^5~M_{\odot}/h$. The higher resolution simulations~($N=720$) allow us to probe seed masses down to $\sim10^4~M_{\odot}$. In Figure \ref{seedmass_dependence}, we reduce the seed mass from $10^5~M_{\odot}/h$ to $1.25\times10^4~M_{\odot}/h$ and study its impact on the final BH mass at $z=7$. We again consider two models that have been shown to successfully produce a $\sim10^9~M_{\odot}/h$ at $z\sim6$. We start with the \texttt{SM5_FOF3000} model~(left panels) where there is substantial amount of merger-driven BH growth. Here we see that the $1.25\times10^4~M_{\odot}/h$ seeds form $\sim8$ times more abundantly compared to $1\times10^5~M_{\odot}/h$ seeds. As a result, $1.25\times10^4~M_{\odot}/h$ seeds undergo more mergers and grow to similar masses as $1\times10^5~M_{\odot}/h$ seeds by $z\sim7$~(as also seen in \citealt{2021MNRAS.507.2012B}). At $z=7$, both the seed masses assemble a $\sim10^8~M_{\odot}$ BH.

Next, we consider the model \texttt{SM5_FOF3000_LW300} which adds a LW flux criterion with $J_{\mathrm{crit}}=300~J_{21}$. This model~(as seen in the previous section) is so restrictive that only a handful of seeds are formed in the entire simulation box~(Figure \ref{seedmass_dependence}: top right panel). Here we allow for super-Eddington accretion with $f_{\mathrm{edd}}=10$~(see Figure \ref{seedmass_dependence}: bottom right panel). We can see that for $M_{\mathrm{seed}}=10^5~M_{\odot}/h$, there are no mergers in its history; and for $M_{\mathrm{seed}}=1.25\times10^4~M_{\odot}/h$, there is only one merger. In the absence of mergers, there is no appreciable growth of these seeds at $z\gtrsim9$. Despite that, both $10^5~M_{\odot}/h$ and $1.25\times10^4~M_{\odot}/h$ seeds grow to $\sim10^8~M_{\odot}/h$ via accretion between $z\sim9-7$. 

To summarize, whether we consider models where $z\sim6$ quasars are formed either with the help of BH mergers, or gas accretion alone, the mass assembled at $z\sim6-9$ is not sensitive to the seed mass between $10^4-10^6~M_{\odot}/h$. Note however that we are not able to probe seed masses below  $\sim10^4~M_{\odot}$ due to resolution limitations. It is possible that these lowest mass seeds~($\sim10^2-10^3~M_{\odot}$) may not be able to grow into the $z\gtrsim6$ quasars, particularly in the absence or lack of mergers;  we shall investigate this in future studies.  

\subsection{Comparison with other theoretical works}
\label{Comparison with other theoretical works}

\begin{figure}
\includegraphics[width=8 cm]{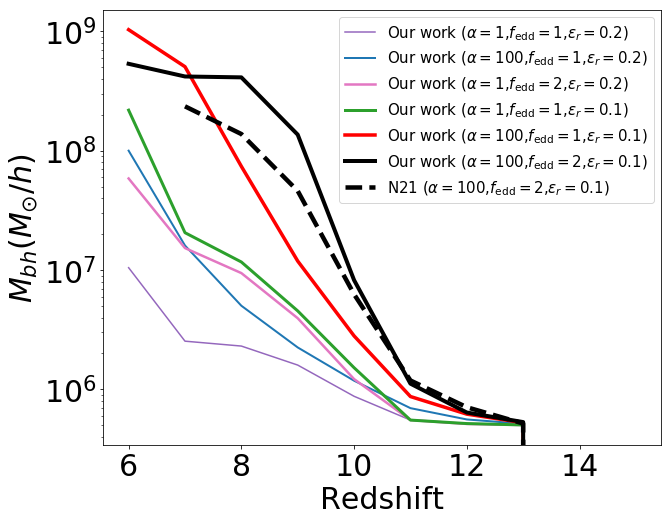}
\caption{Comparison between the BH growth histories for the constrained runs in our work vs \protect\cite{2021MNRAS.tmp.2867N}~(hereafter N21). Their simulations were run using the \texttt{MP-GADGET} code~\citep{yu_feng_2018_1451799} with the galaxy formation model adopted from the \texttt{BlueTides} simulation~\citep{2016MNRAS.455.2778F}. Solid lines show predictions from our simulations, and the dashed line is the prediction from N21. To make a fair comparison, we tune our simulation box, initial conditions to be the same as that used in N21~(box size of $20~\mathrm{Mpc}/h$ with $N=352$). We also use the same ``halo mass based" seeding prescription as N21, with $5\times10^5~M_{\odot}/h$ seeds placed inside $>5\times10^{10}~M_{\odot}/h$ halos. Our fiducial model~(with $\alpha=1,f_{\mathrm{edd}}=1,\epsilon_r=0.2$) assembles a significantly lower BH mass~(by factors of $\sim30$) compared to N21. This difference is due to the the combined effect of the higher Bondi boost factor and Eddington factor, as well as lower radiative efficiency used in N21. In fact, when we use the same accretion parameters as N21~($\alpha=100,f_{\mathrm{edd}}=2,\epsilon_r=0.1$), we produce a slightly higher~(by factor of $\sim2$) mass BH compared to their work.}
\label{arepo_vs_gadget_fig}
\end{figure}

Here, we compare our results to other theoretical works that have explored the formation of the $z\gtrsim6$ quasars. We will first compare with hydrodynamic simulations, where we note that most of the existing work has so far used seed models that are only based on halo mass. Our simulations using the TNG seed model therefore provide the most direct comparison to such studies.  
We start with the constrained simulations of N21 produced using the \texttt{MP-GADGET} code~\citep{yu_feng_2018_1451799} with the \texttt{BlueTides} galaxy formation model~\citep{2016MNRAS.455.2778F}. Their primary constrained peak~(referred to as \texttt{BIG-BH} in their work) is very similar to \texttt{5SIGMA_COMPACT}. Their seed model is also similar to the TNG seed model; they adopt the same halo mass threshold for seeding~($>5\times10^{10}~M_{\odot}/h$), but with a slightly smaller seed mass of $5\times10^5~M_{\odot}/h$. Notably, their simulation produced a $\sim3\times10^8~M_{\odot}$ BH by $z=7$, which is significantly higher than the predictions in our simulations with the TNG seeding and accretion model. But N21 uses a lower radiative efficiency of $\epsilon_r=0.1$ and a Bondi boost factor of $\alpha=100$, which we have already shown to produce a much stronger growth compared to the TNG accretion model~(revisit Figure \ref{varying_accretion_fig}: right panel). Additionally, they also have a higher Eddington factor of $f_{\mathrm{edd}}=2$. 

We perform a more direct comparison to N21 in Figure \ref{arepo_vs_gadget_fig} by simulating a box identical to their work, particularly in terms of volume~($20~\mathrm{Mpc}/h$ box length), resolution~($N=352$), initial condition~(\texttt{BIG-BH}) and the BH seed model~($5\times10^5~M_{\odot}/h$ seeds in $>5\times10^{10}~M_{\odot}/h$ halos). If we apply the TNG accretion model~($\alpha=1,f_{\mathrm{edd}}=1,\epsilon_r=0.2$), our \texttt{BIG-BH} simulation assembles a BH of mass $\sim10^7~M_{\odot}$ by $z=7$~(Figure \ref{arepo_vs_gadget_fig}: purple line); this is $\sim30$ times smaller than the N21 predictions~(similar to that of \texttt{5SIGMA_COMPACT}). Next, if we individually adjust each of these accretion parameters~(Figure \ref{arepo_vs_gadget_fig}: pink, green and blue lines) to the N21 values~(one parameter at a time), we find that they all lead to notable enhancement in the BH growth~(as also seen in Section \ref{Impact of BH accretion sec} for \texttt{SM1000_FOF3000}). Finally, if all the accretion parameters in our simulations are simultaneously set to be the same as N21~(Figure \ref{arepo_vs_gadget_fig}: black line), we produce a $\sim5\times10^8~M_{\odot}$ BH by $z=7$, which is only slightly higher than the N21 predictions. Overall, our results are broadly consistent with N21. The same general conclusion also applies to the comparison with the results of \cite{2020MNRAS.496....1H}, which performs constrained runs using \texttt{MP-GADGET} similar to that of N21. Notably, they find that the final mass at $z\sim6$ is insensitive to the seed mass for $\sim 5\times10^4-5\times10^5~M_{\odot}/h$ seeds, which is consistent with our findings. 

The zoom simulations of \cite{2009MNRAS.400..100S}, \cite{2014MNRAS.440.1865F} and \cite{2014MNRAS.439.2146C} adopted a Bondi boost factor of 100 and radiative efficiencies of 0.05-0.1; with this accretion model, they successfully produced the $z\sim6$ quasars without the need for mergers or super-Eddington accretion, consistent with our results. \cite{2020arXiv201201458Z} performed zoom simulations targeting the formation of a $\sim10^{13}~M_{\odot}$ halo at $z=6$. In their fiducial model, they placed $10^5~M_{\odot}/h$ seeds in $10^{10}~M_{\odot}/h$ halos. These seeds grew via Eddington limited Bondi accretion~($\alpha=1,f_{\mathrm{edd}}=1$) with radiative efficiency of $\epsilon_r=0.1$. With this model, they were able to grow a $\sim10^9~M_{\odot}$ BH by $z=6$ without the help of substantial merger-driven BH growth, super-Eddington accretion, or a Bondi boost. They achieve somewhat faster BH growth compared to our simulations, which only assembles a  $\sim2\times10^8~M_{\odot}$ BH by $z=6$ if $\epsilon_r=0.1$ is applied without a Bondi boost~(see green line in Figure \ref{arepo_vs_gadget_fig}). While it is not clear what may be causing the difference between our results and \cite{2020arXiv201201458Z}, it may be attributed to differences in the implementation of other aspects of the galaxy formation model such as metal enrichment and stellar feedback. However, when they increased their radiative efficiency to 0.2, their final BH mass at $z=6$ dramatically decreased to $\sim4\times10^7~M_{\odot}$, consistent with our findings. Finally, early work by \cite{2007ApJ...665..187L} also  produced a $\sim10^{9}~M_{\odot}$ quasar with $\epsilon_r=0.1$ and $\alpha=1,f_{\mathrm{edd}}=1$ within a $8\times10^{12}~M_{\odot}$ halo at $z=6.5$. Notably, their host halo was assembled after a series of 8 major mergers between $z\sim14-6.5$. Their results therefore highlighted yet another formation pathway for high-z quasars i.e. via a series of intermittent rapid growths driven by major mergers.

Similar to \cite{2020arXiv201201458Z}, \cite{2019MNRAS.488.4004L} was also able to produce a $\sim10^9~M_{\odot}$ by $z\sim6$ with Eddington limited Bondi accretion and radiative efficiency of 0.1, without applying a Bondi boost factor. This may be due to their adopted thermal feedback efficiency of 0.005, which is significantly smaller than values~($\sim0.05-0.15$) adopted for most other works including ours. The very recent work of \cite{2021MNRAS.507....1V} considered even lower radiative efficiencies of $0.03$, allowing them to  grow BHs to $\sim10^9~M_{\odot}$ by $z\sim6$ via feedback regulated Bondi accretion without the need of a Bondi boost factor. Lastly, Radiation hydrodynamics zoom simulations of \cite{2018ApJ...865..126S} also adopted a radiative efficiency of 0.1 and is able to assemble a $\sim10^9~M_{\odot}$ BH by $z\sim7$ despite gas accretion being sub-Eddington~(and no substantial contribution from BH mergers) for almost the entire growth history; however it is difficult to compare their results to our work since they adopted the alpha disk formalism ~(see Eq. (2) of \citealt{2010MNRAS.406L..55D}) to calculate the BH accretion rate, where there is no explicit dependence on BH mass.   

While previous hydrodynamic simulations probing $z\gtrsim6$ quasars have mostly adopted halo mass based prescriptions for seeding, SAMs have been able to explore a broader range of seeding channels~(and seed masses) with more physically motivated seeding criterias~\citep[for example,][]{2007MNRAS.377.1711S,Volonteri_2009, 2012MNRAS.423.2533B,2018MNRAS.476..407V, 2018MNRAS.481.3278R, 2019MNRAS.486.2336D, 2020MNRAS.491.4973D}. Here we shall compare with works that have used SAMs to make predictions specific to the  $z\gtrsim6$ quasars. \cite{2016MNRAS.457.3356V} and \cite{2021MNRAS.506..613S} used the \texttt{GAMETE/QSOdust} data constrained SAM~(introduced in \citealt{2016MNRAS.457.3356V}) to trace the formation of a $\sim10^9~M_{\odot}$ BH along the merger tree of a $10^{13}~M_{\odot}/h$ halo at $z=6.42$. They find that heavy seeds~($\sim10^{5}~M_{\odot}/h$) contribute the most to the formation of $z\gtrsim6$ quasars compared to light~($\sim10^{2}~M_{\odot}/h$) and intermediate seeds~($\sim10^{3}~M_{\odot}/h$). While we cannot probe light and intermediate seeds, their results for the heavy seeds do not conflict with our findings.  They also apply a radiative efficiency of 0.1 combined with a Bondi boost factor of $50-150$, and are able to assemble $z\sim6$ quasars without substantial contributions from BH mergers or super-Eddington accretion. This is fully consistent with our results, and also with results from other hydrodynamic simulations. 

Overall, we find that the differences between our results with \texttt{IllustrisTNG} physics and the results from most previous works, are largely originating from differences in the modeling of BH accretion and feedback. This also brings to light that when the default \texttt{IllustrisTNG} physics is applied to such extreme overdense regions, it is much more difficult to form $z\sim6$ quasars compared to the physics adopted in other simulations and SAMs. This is primarily due to the TNG accretion model which has a higher radiative efficiency of 0.2~(most works adopt a value of 0.1). At the same time, the lack of a Bondi boost also slows the BH growth even further~(most works adopted a value of 100) in the TNG accretion model. Note that the uncertainties within the modeling of BH accretion are significant, particularly at high redshifts wherein the gas environments are likely to be very different compared to the assumptions underlying the Bondi accretion model. Additionally, the radiative efficiencies are also poorly constrained. To that end, different subgrid models are better or worse at reproducing different aspects of the observed SMBH and galaxy populations~\citep[e.g.][]{2021MNRAS.503.1940H,2022MNRAS.509.3015H}. Moving forward, it will be necessary to build better subgrid models with fewer modeling uncertainties, as well as improving the observational constraints particularly on high-z SMBHs.

\section{Discussion and Conclusions}
\label{Conclusions_sec}
In this work, we have studied the implications of the \texttt{IllustrisTNG} galaxy formation model on the brightest $z\gtrsim6$ quasar population, particularly in the context of different BH seeding models. These extremely rare~($\sim1~\mathrm{Gpc}^{-3}$) objects have grown to masses of $\sim10^9-10^{10}~M_{\odot}$~(comparable to the most massive $z\sim0$ SMBHs) within the first Gyr since the Big Bang; this is difficult to achieve in general, and it is likely to place strong constraints on models for BH formation and growth. 

We explore the following seeding prescriptions:
\begin{itemize}
    \item TNG seed model: This is the default ``halo mass based" prescription used within the \texttt{IllustrisTNG} simulation suite, where we place $8\times10^5~M_{\odot}/h$ seeds in $>5\times10^{10}~M_{\odot}/h$ halos. 
    \item gas-based seed models: Here we place seeds~($M_{\mathrm{seed}}=1.25\times10^4,1\times10^5~\&~8\times10^5~M_{\odot}/h$) in halos that exceed critical thresholds for halo mass and dense, metal poor gas mass~(represented by $\tilde{M}_h$ and $\tilde{M}_{\mathrm{sf,mp}}$ respectively in the units of $M_{\mathrm{seed}}$). We also explore models where the dense, metal poor gas is required to have LW fluxes above a critical value $J_{\mathrm{crit}}$. 
\end{itemize}


With the above seeding prescriptions, we probe the possible formation of the $z\sim6$ quasars~(defined as $\sim10^9~M_{\odot}$ BHs with luminosities of $\sim10^{47}~\mathrm{erg~s^{-1}}$) within extremely rare peaks in the density field using the technique of constrained Gaussian realizations. This technique allows us to constrain the peak of the density field so as to assemble $\gtrsim10^{12}~M_{\odot}/h$ halos by $z\sim7$ within a simulation volume of $(9~\mathrm{Mpc/h})^3$. Having a relatively small simulation volume allows us to build a large simulation suite exploring a variety of density peak parameters as well as seeding parameters.    

We reproduce findings from previous work~(N21) showing that BH growth is most efficient at density peaks that have high compactness and a low tidal field. In fact, a highly compact $5\sigma$ peak at $1.0$ Mpc/h with low tidal field~(\texttt{5SIGMA_COMPACT}) produces a more massive BH~(by factors of $\sim2$) compared to a typical $6\sigma$ peak at $1.3$ Mpc/h~(\texttt{6SIGMA}). The reason for this is two-fold: First, the target $z=7$ halo in \texttt{5SIGMA_COMPACT} has more massive progenitors than that of \texttt{6SIGMA}, allowing seeds to form in potentially higher numbers and boosting the merger-driven BH growth. Second, the \texttt{5SIGMA_COMPACT} region forms a more compact gas cloud which falls towards the BH more symmetrically from all directions compared to \texttt{6SIGMA}; this leads to higher gas densities in their neighborhood and boosts the accretion-driven BH growth. 

Despite the enhanced accretion and merger-driven BH growth in \texttt{5SIGMA_COMPACT}, we find that when the TNG seed model is used, the final mass of the central BH in the target halo at $z=6$ is only $\sim5\times10^7~M_{\odot}$ with luminosities of $\sim10^{45}~\mathrm{erg~s^{-1}}$. This significantly falls short of producing an observed $z\sim6$ quasar i.e. a $\sim10^9~M_{\odot}$ BH with a bolometric luminosity of $\sim10^{47}~\mathrm{erg~s^{-1}}$. But when we apply the more physically motivated gas-based seeding prescription where BHs are seeded in halos with minimum star forming metal poor gas mass of $5$ times the seed mass and a total halo mass of $3000$ times the seed mass~($\tilde{M}_{\mathrm{sf,mp}}=5$ and $\tilde{M}_{\mathrm{h}}=3000$), we find that there is substantial amount of the merger-driven BH growth at $z\gtrsim10$ compared to the TNG seed model. As a result, the BH assembles a mass of $\sim10^{9}~M_{\odot}$ at $z=6$ and grows close to the Eddington limit with a bolometric luminosity of  $\sim10^{47}~\mathrm{erg~s^{-1}}$. This is consistent with the observed $z\sim6$ quasars, and is achievable for all seed mass values between $\sim10^4-10^6~M_{\odot}/h$. Lastly, note that this can also be achieved by enhancing merger-driven growth within halo-mass based seed models~(like TNG seed model) by sufficiently reducing the halo mass threshold.

Notably, there are two distinct phases in the BH growth in our simulations: 1) $z\gtrsim9$ when the BH growth is predominantly driven by BH mergers, and 2) $z\sim9-6$ when gas accretion dominates the BH growth. 
To form a $z\gtrsim6$ quasar within a universe with  \texttt{IllustrisTNG} physics, the BH growth has to be boosted by BH mergers at $z\gtrsim9$. Amongst all the seed models we explored, only the one with $\tilde{M}_{\mathrm{sf,mp}}=5$ and $\tilde{M}_{\mathrm{h}}=3000$ provides enough mergers to assemble $z\sim6$ quasars.

For much more restrictive gas-based seed models~($\tilde{M}_{\mathrm{sf,mp}}=1000$ and $\tilde{M}_{\mathrm{h}}=3000$, for example), very few seeds are formed and there is little to no merger-driven growth; as a result, they fail to produce $z\sim6$ quasars in the \texttt{IllustrisTNG} universe. However, recall that the \texttt{IllustrisTNG} model was calibrated to reproduce properties of relatively common galaxies and BHs at low redshifts. We explored the possibility of enhanced accretion in these extreme overdense regions compared to the TNG accretion model. We found that in order to form $z\sim6$ quasars with these restrictive seed models, it is crucial to increase the maximum accretion rate~(by factors $\gtrsim10$) allowed for a BH of a given mass to grow. This can be achieved by either increasing the Eddington factor or decreasing the radiative efficiency. To that end, increasing the Bondi boost factor alone does not sufficiently boost the BH mass assembly to produce the $z\sim6$ quasars. Lastly, note that even for such high values of  $\tilde{M}_{\mathrm{sf,mp}}$, one can enhance merger-driven BH growth by choosing a lower halo mass threshold $\tilde{M}_{\mathrm{h}}$; this would relax the constraints on the accretion model in producing $z\sim6$ quasars.


Prospects for DCBH formation in the \texttt{5SIGMA_COMPACT} region are limited if the critical LW fluxes are indeed $\gtrsim1000~J_{21}$ as predicted by one-zone chemistry models and small scale hydrodynamics simulations. This is because LW intensities within the dense, metal poor pockets of \texttt{5SIGMA_COMPACT} region do not significantly exceed $\sim300~J_{21}$ between $z\sim7-22$. \texttt{5SIGMA_COMPACT} region produces a handful of seeds for somewhat lower critical fluxes, particularly $\sim300~J_{21}$. Even for these optimistic estimates of $J_{\mathrm{crit}}$, due to the obvious lack of merger-driven BH growth, DCBHs would require one of the optimal accretion scenarios described in the previous paragraph in order to grow a $z\gtrsim6$ quasar. As far as other theoretical seeding channels such as Pop III and NSC seeds, without being able to explicitly resolve their formation conditions, it is currently difficult to tell whether they form and merge abundantly enough to qualify as potential origins of the $z\gtrsim6$ quasars; we shall investigate this in the future.  

We note that our results are specific to features of our underlying  \texttt{IllustrisTNG} galaxy formation model. They may significantly depend on the prescriptions for star formation, metal enrichment, stellar feedback and BH dynamics. Additionally, there are also several other BH seeding, accretion  and feedback models beyond the ones explored in this work, that could potentially produce $z\sim6$ quasars. Black hole accretion and feedback is a major source of uncertainty. For example, the lack of accretion-driven BH growth at $z\gtrsim9$ may be partly influenced by the Bondi accretion model which struggles to grow low mass BHs due to the $M_{bh}^2$ scaling of the accretion rate. This $M_{bh}^2$ scaling also implies that at these early epochs when the self-regulation by feedback is relatively weak, the BH growth would be extremely sensitive to the local gas environment. This local gas environment may be impacted by other aspects of galaxy formation, such as star formation, stellar feedback~\citep[for e.g.][]{2017MNRAS.468.3935H}, metal enrichment and gas cooling. While the $M_{bh}^2$ scaling appears as a generic feature of all accretion models based on a gas capture radius 
\citep{2005MNRAS.361..776S,2007ApJ...665..107P,2009MNRAS.398...53B}, there are also models such as gravitational torque driven accretion~\citep{2017MNRAS.464.2840A,2019MNRAS.486.2827D} where the scaling exponent is smaller~($M_{bh}^{1/6}$). This can significantly boost the growth of low mass BHs, but also slow down the growth of high mass BHs. As a result, it can have non-trivial implications for the feasibility of various BH models to produce $z\gtrsim6$ quasars. 

A final caveat to our results lies within our modelling of BH dynamics. In particular, due to the limited simulation resolution, we use the standard BH repositioning scheme which instantaneously relocates the BH to a nearby potential minimum. In fact, several simulations with more realistic dynamics models~\cite[e.g.][]{2017MNRAS.470.1121T} have now indicated that it may be difficult for many of the seeds~(particularly lower mass seeds) to sink to the local potential minima, thereby leading to a population of wandering BHs~\citep{2018ApJ...857L..22T,2021MNRAS.503.6098R,2021ApJ...916L..18R,2021MNRAS.508.1973M,2022MNRAS.tmp..221W}. This would have two important effects: 1) over-estimating the accretion rates since the BHs may spend more time around dense gas compared to more realistic dynamics models, and 2) nearby BHs are promptly merged, thereby overestimating the merger rates at early times.  In the future, we shall assess the impact of all of these caveats on the formation of $z\gtrsim6$ quasars.

Despite the caveats, our results overall indicate a strong prospect of revealing the seeding environments for the observed $z\gtrsim6$ quasars using upcoming facilities such as LISA. In particular, regardless of the accretion model, different seed models predict distinct merger and accretion histories for the progenitors of these quasars at $z\gtrsim9$. These progenitors will also be amongst the most massive sources at their corresponding redshift. In addition to the strong prospect of detecting their mergers with LISA up to $z\sim20$, their AGN luminosities also exceed detection limits of Lynx and JWST up to $z\sim10$. However, the difference in the predicted AGN luminosities between different seed models is small~($\lesssim2.5$ dex in magnitude). Therefore, detecting electromagnetic signatures of seeding is going to be challenging for JWST and Lynx.

\section*{Acknowledgements}
AKB thanks Dylan Nelson for valuable discussion and feedback. LB acknowledges support from NSF award AST-1909933 and Cottrell Scholar Award \#27553 from the Research Corporation for Science Advancement.
PT acknowledges support from NSF-AST 2008490.
RW is supported by the Natural Sciences and Engineering Research Council of Canada (NSERC), funding reference \#CITA 490888-16.
TDM acknowledges funding from NSF AST-1616168, NASA ATP  80NSSC20K0519,
 NASA ATP 80NSSC18K101, and NASA ATP NNX17AK56G.
This work was also supported by the NSF AI Institute: Physics of the Future, NSF PHY-2020295. YN acknowledges support from the McWilliams fellowship.
\section*{Data availablity}
The underlying data used in this work shall be made available upon reasonable request to the corresponding author.



\bibliography{references}
\end{document}